\documentclass[10pt,letterpaper]{article}
\usepackage[top=0.85in,left=2.75in,footskip=0.75in]{geometry}

\usepackage{amsmath,amssymb}
\usepackage{changepage}
\usepackage[utf8x]{inputenc}
\usepackage{textcomp,marvosym}
\usepackage{cite}
\usepackage{ulem}
\usepackage{nameref,hyperref}
\usepackage{microtype}
\usepackage{nicefrac}
\DisableLigatures[f]{encoding = *, family = * }
\usepackage[table]{xcolor}
\usepackage{array}
\newcolumntype{+}{!{\vrule width 2pt}}
\newlength\savedwidth

\raggedright
\usepackage[aboveskip=1pt,labelfont=bf,labelsep=period,justification=raggedright,singlelinecheck=off]{caption}

\makeatletter
\renewcommand{\@biblabel}[1]{\quad#1.}
\makeatother

\usepackage{lastpage,fancyhdr,graphicx}
\usepackage{epstopdf}
\fancyheadoffset[L]{2.25in}
\fancyfootoffset[L]{2.25in}
\usepackage{bm}
\usepackage{todonotes}

\newcommand{\new}[1]{#1}
\newcommand{\old}[1]{}

\begin{document}
\vspace*{0.2in}
\normalem
\begin{flushleft}
{\Large
\textbf\newline{Topological data analysis distinguishes parameter regimes in the Anderson-Chaplain model of angiogenesis} % Please use "sentence case" for title and headings (capitalize only the first word in a title (or heading), the first word in a subtitle (or subheading), and any proper nouns).
}
\newline
\\
John T. Nardini\textsuperscript{1},
Bernadette J. Stolz\textsuperscript{2},
Kevin B. Flores\textsuperscript{1},
Heather A. Harrington\textsuperscript{2,3},
Helen M. Byrne*\textsuperscript{2}
\\
\bigskip
\textbf{1} Department of Mathematics, North Carolina State University, Raleigh, North Carolina, USA
\\
\textbf{2} Mathematical Institute, University of Oxford, Oxford, OX2 6GG, UK
\\
\textbf{3} Wellcome Centre for Human Genetics, University of Oxford, Oxford, OX3 7BN, UK
\\
\bigskip

% Use the asterisk to denote corresponding authorship and provide email address in note below.
* helen.byrne@maths.ox.ac.uk
\end{flushleft}

%\todo[inline,backgroundcolor=OliveGreen!25]{Words not to use: \\ -Reliable (I have now changed this to ``robust", but am open to other suggestions)\\ - network inference\\ - any instance of ``network" should say ``vessel network" or ``simulated vessel network". \\  
%-"Standard biological descriptors" was changed to "standard descriptors," except the first mention is "morphologically-motivated standard descriptors" 
%}

\section*{Abstract}
Angiogenesis is the process by which blood vessels form from pre-existing vessels. It plays a key role in many biological processes, including embryonic development and wound healing, and contributes to many diseases including cancer and rheumatoid arthritis. The structure of the resulting vessel networks determines their ability to deliver nutrients and remove waste products from biological tissues. 
Here we simulate the Anderson-Chaplain model of angiogenesis at different parameter values and quantify the vessel architectures of the resulting synthetic data. Specifically, we propose a topological data analysis (TDA) pipeline for systematic analysis of the model. TDA is a vibrant and relatively new field of computational mathematics for studying the shape of data. We compute topological and standard descriptors of model simulations generated by different parameter values. We show that TDA of model simulation data stratifies parameter space into regions with similar vessel morphology. The methodologies proposed here are widely applicable to other synthetic and experimental data including wound healing, development, and plant biology.

% Please keep the Author Summary between 150 and 200 words
% Use first person. PLOS ONE authors please skip this step. 
% Author Summary not valid for PLOS ONE submissions.   
\section*{Author summary}
 Vascular networks play a key role in many physiological processes, by delivering nutrition to, and removing waste from, biological tissues. In cancer, tumors stimulate the growth of new blood vessels, via a process called angiogenesis. The resulting vascular structure comprises many inter-connected vessels which lead to emergent morphologies that influence the rate of tumor growth and treatment efficacy. In this work, we consider several approaches to summarize the morphology of synthetic vascular networks generated from a mathematical model of tumor-induced angiogenesis. We find that a topological approach can be used quantify vascular morphology of model simulations
 and group the simulations into biologically interpretable clusters. This methodology  may be useful for the diagnosis of abnormal blood vessel networks and quantifying the efficacy of vascular-targeting treatments.

%\linenumbers

% Use "Eq" instead of "Equation" for equation citations.
\section*{Introduction}

% P1 : Spatial structure of networks

Blood vessels deliver nutrients to, and remove waste products from, tissues during many physiological processes, including embryonic development, wound healing, and cancer \cite{gupta_mechanism_2003,kushner_building_2013,tonnesen_angiogenesis_2000}. The structure and form of connected blood vessels (e.g., vascular morphology), determines how nutrients and waste are supplied or removed from the environment and, in turn, influences the behavior of the tissue and its constituent cells. The morphology of vascular networks can reveal the presence of an underlying disease, or predict the response of a patient to treatment.  High resolution vascular imaging technology creates an exciting opportunity to develop mathematical tools to discover new links between blood vessel structure and function.

\begin{figure}
    \centering
    \includegraphics[width=0.57\textwidth]{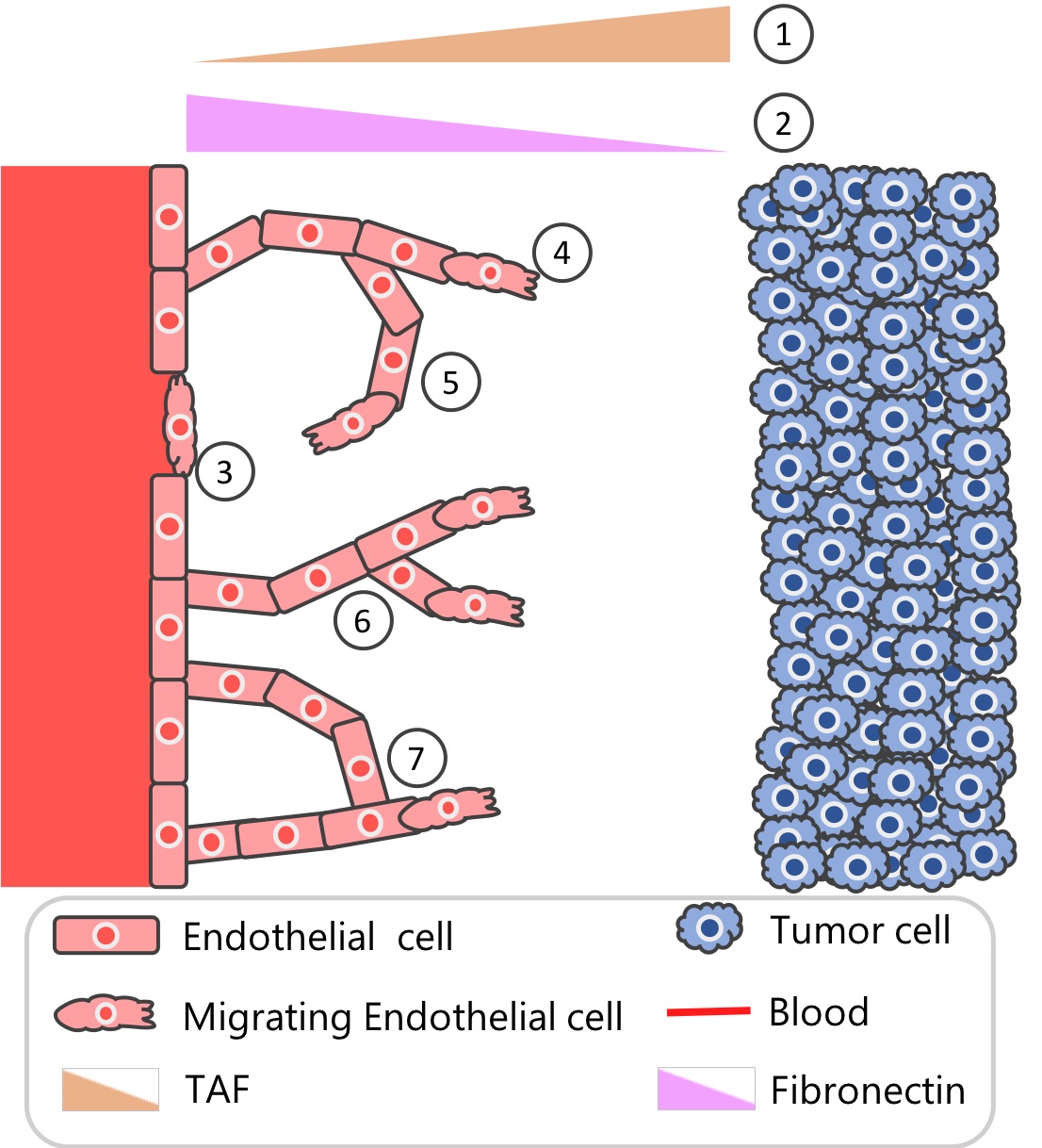}
    \caption{\textbf{Schematic overview of tumor-induced angiogenesis.} \new{Schematic depicting seven aspects of tumor angiogenesis that are incorporated in the Anderson-Chaplain model of tumor-induced angiogenesis \cite{anderson_continuous_1998}. (1) Distant tumor cells release a  range of chemoattractants, including vascular endothelial growth factors and basic fibroblast growth factors that stimulate the formation of new blood vessels. These growth factors are described collectively as a single, generic tumor angiogenesis factor (TAF). (2) Production and consumption of tissue-matrix bound fibronectin by the endothelial cells, creates a spatial gradient of fibronectin across the domain. (3) Endothelial cells sense the TAF and fibronectin gradients and undergo individual cell migration. (4) As endothelial tip cells migrate, via chemotaxis, up spatial gradients of the TAF, stalk cells in the developing vessels are assumed to proliferate, creating  what has been termed a ``snail trail" of new endothelial cells. (5) Endothelial tip cells also migrate, via haptotaxis, up spatial gradients in fibronectin. (6) Endothelial cells in existing sprouts may initiate the formation of secondary sprouts. (7) If a sprout coincides with an existing vessel, then it is assumed to be annihilated and a new loop is formed; if two sprouts coincide or anastomose, then both are assumed to be annihilated and a new loop forms.}
%\caption{
\old{ Schematic of \new{the processes included in the Anderson-Chaplain Model of }tumor-induced angiogenesis. We depict seven  aspects of tumor-induced angiogenesis that are captured in the Anderson-Chaplain model \cite{anderson_continuous_1998}. (1) Tumor cells release a chemoattractant, or tumor angiogenesis factor (TAF), that stimulates the formation of new blood vessels. (2) \old{Endothelial cells lining pre-existing blood vessels} \new{In response to the presence of TAF, the parent vessel becomes permeable and releases plasma fibronectin. This plasma fibronectin then binds to the substrate, which creates a gradient of fibronectin. (3) In response to TAF, endothelial cells may sense the TAF gradient and exhibit individual cell migration}. (4) Endothelial tip cells migrate, via chemotaxis, up spatial gradients of the TAF and deposit a ``snail trail" of new vessels \new{comprised of endothelial cells} as they move. (5) Endothelial tip cells also migrate, via haptotaxis, up spatial gradients in fibronectin which is bound to the tissue matrix. (6) A migrating endothelial cell can branch into two cells, to form a new sprout. (7) If a sprout coincides with an existing vessel, then it is annihilated and a new loop is formed; if two sprouts coincide, or anastomose, then they are both annihilated and a new loop forms.}}
    \label{fig:fig1}
\end{figure}

\new{Here, we focus on \emph{tumor-induced angiogenesis}, the process by which tumor cells stimulate the formation of new blood vessels from pre-existing vasculature \cite{gupta_mechanism_2003}. When oxygen and nutrient levels within a population of tumor cells become too low to sustain a viable cell population, the tumor cells produce several growth factors, including vascular endothelial growth factor (VEGF), platelet-derived growth factor (PDGF) and basic fibroblast growth factor (bFGF), which diffuse through the surrounding tissue \cite{ferrara_vegf_2002,folkman_tumor_1974,sherwood_tumor_1971}. On contact with neighboring blood vessels, these tumor angiogenesis factors (or TAFs) increase the permeability of the vessel walls and loosen adhesive bonds between the endothelial cells that line the blood vessels \cite{folkman_isolation_1971}. 
The TAFs activate the endothelial cells to release proteases that degrade the basement membrane \cite{menashi_2_1993}. Activated tip endothelial cells then migrate away from the parent vessel and follow external cues, such as spatial gradients of VEGF and/or fibronectin \cite{paweletz_tumor-related_1989,terranova_human_1985}. Stalk endothelial cells located behind the tip cells proliferate into the surrounding tissue matrix, causing the emerging vessel sprouts to elongate \cite{sholley_mechanisms_1984}.  When tip endothelial cells from one sprout come into contact with tip or stalk endothelial cells from another sprout, the two endothelial cells may fuse or anastamose together, forming a new loop through which oxygen and nutrients may be transported \cite{paweletz_tumor-related_1989}. In Fig~\ref{fig:fig1} we present a schematic model of how these processes are coordinated. }
 
\new{As the number of experimental and theoretical studies of angiogenesis has increased, so our knowledge of the mechanisms involved in its regulation has increased beyond the idealised description given above \cite{lugano_tumor_2020,saman_inducing_2020}. 
For example, subcellular signalling involving the VEGF and delta-notch signalling pathways is known to influence whether an endothelial cell adopts a tip or stalk cell phenotype \cite{bentley_role_2014}. Further, immune cells, including macrophages, that are present in the tumor microenvironment, are known to release TAFs such as VEGF and matrix degrading proteases which facilitate tumor angiogenesis \cite{lugano_tumor_2020}. Mechanical stimuli also influence endothelial cell migration. For example, the fluid shear stress experienced by endothelial cells lining blood vessels drives their migration by causing them to form lamellipodia and focal adhesions at the front of the cell, in the flow direction, and to retract focPrincipal component analysis of persistent homology rank functions with case studies of spatial point patterns, sphere packing and colloidal adhesions at the rear \cite{li_notch_2005,li_role_2002}. }

\new {In this paper}, we explore several approaches for quantifying the morphology of vascular networks that form during angiogenesis, including recently developed techniques from topological data analysis. We develop and test our methodology through application to simulated data from a
highly idealised, and well known hybrid (i.e., discrete and continuous) stochastic mathematical model of tumor-induced angiogenesis proposed by Anderson and Chaplain \cite{anderson_continuous_1998} (see Fig~\ref{fig:fig1} for a model schematic).  

The Anderson-Chaplain model of tumor angiogenesis is based on continuum models proposed by Balding and McElwain \cite{balding_mathematical_1985} \new{and Byrne and Chaplain
\cite{byrne_growth_1995}}, as well as the stochastic model proposed by Stokes and Lauffenberger \cite{stokes_analysis_1991}. It is an on-lattice model in which endothelial cells perform a biased random walk and generate blood vessel networks that resemble those observed experimentally \cite{anderson_continuous_1998}. While more detailed theoretical models have been developed (see reviews \cite{byrne_dissecting_2010,hadjicharalambous_tumour_2021,metzcar_review_2019,scianna_review_2013,stepanova_multiscale_2021}), we focus on the Anderson-Chaplain model due to its simplicity 
\new{and wide adoption in the mathematical biology literature. The methods presented in this work can also be used to study alternative models of angiogenesis, such as the phase-field model presented in \cite{vilanova_mathematical_2017}, the 2D model of early angiogenesis and cell fate specification in \cite{stepanova_multiscale_2021}, the 3D hybrid model presented in \cite{perfahl_3d_2017}, and multiscale models of vascular tumor growth, such as those presented in \cite{grogan_microvessel_2017, vavourakis_validated_2017,cai_long_2017,sefidgar_numerical_2015}.} 

The \new{Anderson-Chaplain} model describes the spatio-temporal evolution of three physical variables: endothelial tip cells, tumor angiogenesis factor (TAF), and fibronectin \cite{anderson_continuous_1998}. Fig~\ref{fig:fig1} contains a model schematic where tumor cells located along the right boundary produce TAF and the endothelial tip cells from a nearby healthy vessel placed along the left boundary move via \emph{chemotaxis} up spatial gradients of TAF . \new{As the tip cells move, stalk cells located behind them proliferate, creating what has been termed} \old{they leave behind} a ``snail trail''  of new blood vessel segments \cite{byrne_dissecting_2010}.
\new{Endothelial tip cells also migrate via \emph{haptotaxis}, up spatial gradients of fibronectin, which is produced and consumed by the migrating endothelial cells and binds to the tissue matrix in which the cells are embedded. 
%We assume that a line of tumor cells at the rightmost boundary of the domain, which creates a spatial TAF gradient. Endothelial tip cells that begin at the left-most boundary of the domain chemotax, or migrate towards increasing TAF gradient, with rate $\chi$ which leads them towards the tumor area. 
As the tip cells migrate and the stalk cells proliferate, the vessel sprouts elongate, and secondary sprouts may emerge from the vessels, at the end of a sprout or laterally from the nascent vessel. Further, when a tip cell collides with another vessel segment or another tip cell, the two cells may fuse, forming a closed loop or anastomosis}.  

\begin{figure}[ht!]
    \centering
    \includegraphics[width=.99\textwidth]{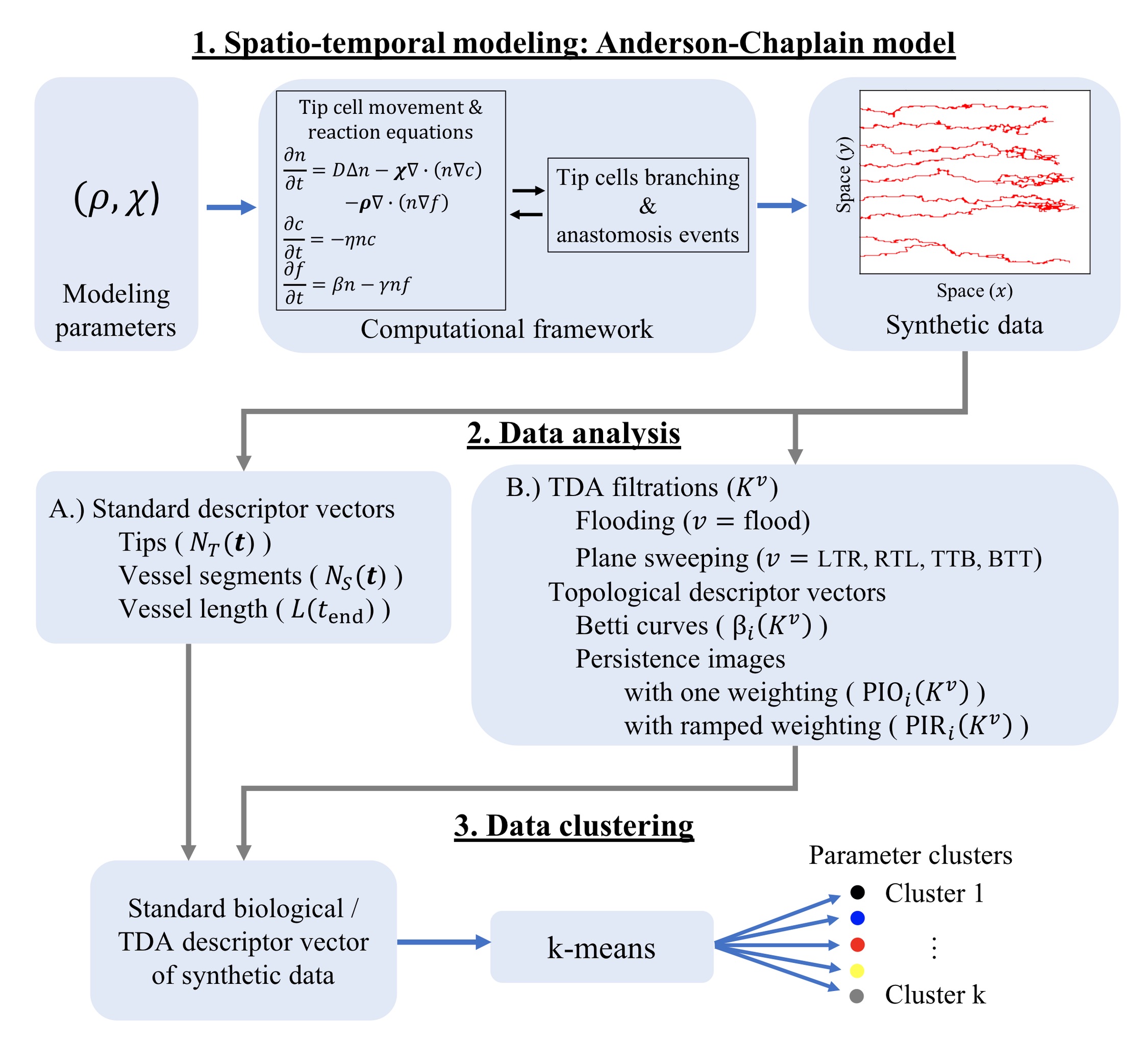}
    \caption{\textbf{Data generation and analysis pipeline.}
(1) \uline{Spatio-temporal modeling: Anderson-Chaplain model.} The Anderson-Chaplain model is simulated for 11 $\times$ 11 = 121 different values of the haptotaxis and chemotaxis parameters, ($\rho, \chi$). The model output is saved as a binary image, where pixels labeled 1 or 0 denote the presence or absence, respectively, of blood vessels. We generate 10 realizations for each of the 121 parameter combinations, leading to 1,210 images of simulated vessel networks. (2) \uline{Data analysis.} \new{We use the binary images from Step 1 to generate standard and topological descriptor vectors}
A.) Standard descriptor vectors We compute the number of active tip cells and the number of vessel segments \new{at discrete time points} \old{over time}. We also compute the length of the vessels at the final time point \cite{folarin_three-dimensional_2010,konerding_evidence_1999,konerding_3d_2001,adams_persistence_2017}. B.) Topological descriptor vectors. We construct flooding and sweeping plane filtrations \cite{bendich_persistent_2016} \new{using binary images from the final timepoint of each simulation.} We track the birth and death of topological features (connected components and loops) that are summarized as Betti curves and persistence images \cite{adams_persistence_2017}. (3) \uline{Data clustering.} We \old{next} perform $k$-means clustering \new{using either} the \new{standard or topological} descriptor vectors \old{for the set of 1,210 simulations} computed during step 2 \new{from all 1,210 simulations. In this way, we} \old{to} decompose \old{the} ($\rho,\chi$) parameter space into regions that group vessel networks with similar morphologies.}
    \label{fig:fig2}
\end{figure}

We would like a quantitative method to understand and compare vessel networks that is applicable to experimental and synthetic data. Standard quantitative descriptors of vascular morphology proposed by ~\cite{folarin_three-dimensional_2010,kannan_functional_2018,konerding_evidence_1999,konerding_3d_2001, perfahl_3d_2017} include inter-vessel spacing, the number of branch points, measuring the total area (volume) covered for 2D (3D) simulations, vessel length, density, tortuosity, the number of vessel segments, and the number of endothelial sprouts. These metrics correlate with tumor progression and treatment response in experimental datasets \cite{bauer_cell-based_2007}; however, they are computed at a fixed spatial scale, and neglect information about the connectedness (i.e., the topology) of vascular structures. 
Therefore, standard descriptors do not exploit the full information content of synthetic data generated from the models. At the same time, advances in imaging technology mean that it is now possible to visualize vascular architectures across multiple spatial scales \cite{zhan_mri_2019}. These imaging developments are creating a need for new methods that can
quantify the multi-scale patterns in vascular networks, and how they change over time.

A relatively new and expanding field of computational mathematics for studying the shape of data at multiple scales is TDA. TDA consists of algorithmic methods for identifying global structures in high-dimensional datasets that may be noisy \cite{carlsson_topology_2009,zomorodian_topology_2005}. TDA has been useful for biomedical applications \cite{nicolau_topology_2011,nielson_topological_2015,qaiser_persistent_2016,xia_persistent_2014}, including brain vessel MRI data \cite{bendich_persistent_2016} and experimental data of tumor blood vessel networks \cite{Byrne2019,Stolz2020}. A commonly-used tool from TDA is \emph{persistent homology} (PH). \emph{Homology} refers to the characterization of topological features (\emph{e.g.}, connected components and loops) in a dataset, and \emph{persistence} refers to the extent to which these features persist within the data as a
scale parameter varies. PH is a useful way to summarize the topological characteristics of spatial network data \cite{Byrne2019,Stolz2020,feng_persistent_2019} as well as noisy and high dimensional agent-based models (ABMs) \cite{topaz_topological_2015, mcguirl_topological_2020}. More recently, PH has been combined with machine learning to extract accurate estimates of parameters from such models \cite{bhaskar_analyzing_2019}. Most of these studies focus on PH calculations for point clouds of data; notably, PH has been adapted to analyze networks and grayscale images \cite{Otter2017}. An active area of research is exploring different filtrations for application-driven PH.

In this paper, we develop a computational pipeline for analyzing simulations of the Anderson-Chaplain model over a range of biologically-relevant parameter values. Our main objectives are to identify model descriptor vectors that group together \new{similar model simulations into \emph{biologically interpretable} clusters. By biologically interpretable, we mean that simulations within a cluster arise from similar parameter values and simulations in different clusters arise from different parameter values.} \old{similar model simulations, and then to interpret the resulting clusters. }

We use two topological approaches, the sweeping plane and flooding filtrations, to construct simplicial complexes from \new{binary image} data generated from simulations of the Anderson-Chaplain model. We show that PH of the sweeping plane filtration and its subsequent vectorization provides an interpretable descriptor vector for the model parameters governing chemotaxis and haptotaxis. We show, \new{by comparison with existing morphologically-motivated standard descriptor vectors, that this topological approach} leads to more biologically \new{interpretable}\old{informative} clusters that \new{stratify the haptotaxis-chemotaxis parameter space} \old{existing morphologically-motivated standard descriptor vectors}. Furthermore, the clusters generated from the sweeping plane filtration are robust and generalize well to unseen model simulations.

\section*{Materials and Methods}

We simulate the Anderson-Chaplain model of tumor-induced angiogenesis for a range of values of the haptotaxis and chemotaxis coefficients, $\rho$ and $\chi$. Each simulation generates vasculature composed of a collection of interconnected blood vessels in the form of binary image data. \new{We compute three so-called standard descriptor vectors to summarize each simulation: two are computed at 50 time steps and the third is computed at the final time step.} We also compute 30 topological descriptor vectors \old{for each model simulation}. These topological descriptors encode the proposed multiscale quantification of the vascular morphology. We calculate standard and topological descriptor vectors for the entire set of 1,210 simulations. We then analyze the  simulations by clustering the standard or topological descriptor vectors and compare the results.    Fig~\ref{fig:fig2} shows the pipeline of data generation and analysis. We give details of the Anderson-Chaplain model, including governing equations, parameter values and implementation, in the S1 Appendix. 
The Python files and notebooks used for all steps of our analysis are publicly available at  \href{https://github.com/johnnardini/Angio_TDA}{https://github.com/johnnardini/Angio\_TDA}.

\subsection*{Data generation}

We simulate the Anderson-Chaplain model as the parameters $\rho$ and $\chi$ are varied independently among the following 11 values: $\{0,0.05,0.1,0.15,\dots , 0.5\}$. 
\new{The parameter bounds of [0,0.5] were chosen to yield a range of vascular architectures that are visibly distinct and recapitulate
vascular patterns seen \emph{in vivo}. The resolution of the parameter mesh (11 discrete values between 0 and 0.5) was chosen for computational tractability while still being sufficient to illustrate how TDA methods can detect and quantify differences between simulations from similar parameter values that are difficult to distinguish visually.} 
We generate 10 realizations of the model for each of the 121 $(\rho,\chi)$ parameter combinations, and produce 1,210 binary images that summarize the different synthetic blood vessel networks. \old{Each simulation runs until either an endothelial tip cell crosses the line corresponding to $x=0.95$ nondimensional spatial units or the maximum simulation time of $t=20$ nondimensional time units is exceeded. Each simulation is initialized using the same initial condition and with all other model parameters set to the baseline values from \cite{anderson_continuous_1998}. We use no-flux boundary conditions.}\new{Each simulation is initialized using the same initial conditions and with all other model parameters fixed at the baseline values used in \cite{anderson_continuous_1998}, and impose no-flux conditions on the domain boundary. For consistency with \cite{anderson_continuous_1998}, all simulations are computed on lattices of size $201\times 201$. Each simulation continues until either an endothelial tip cell crosses the line $x=0.95$ (in dimensionless spatial units) or the maximum simulation time, $t=20$ (in dimensionless time units), is exceeded. The final model output is a $201\times 201$ binary image in which nonzero (zero) pixels denote the presence (absence) of an endothelial cell at that lattice site.}

\subsubsection*{Morphologically-motivated standard descriptors and descriptor vectors}

\begin{figure}[ht!]
    \centering
    \includegraphics[width=.7\textwidth]{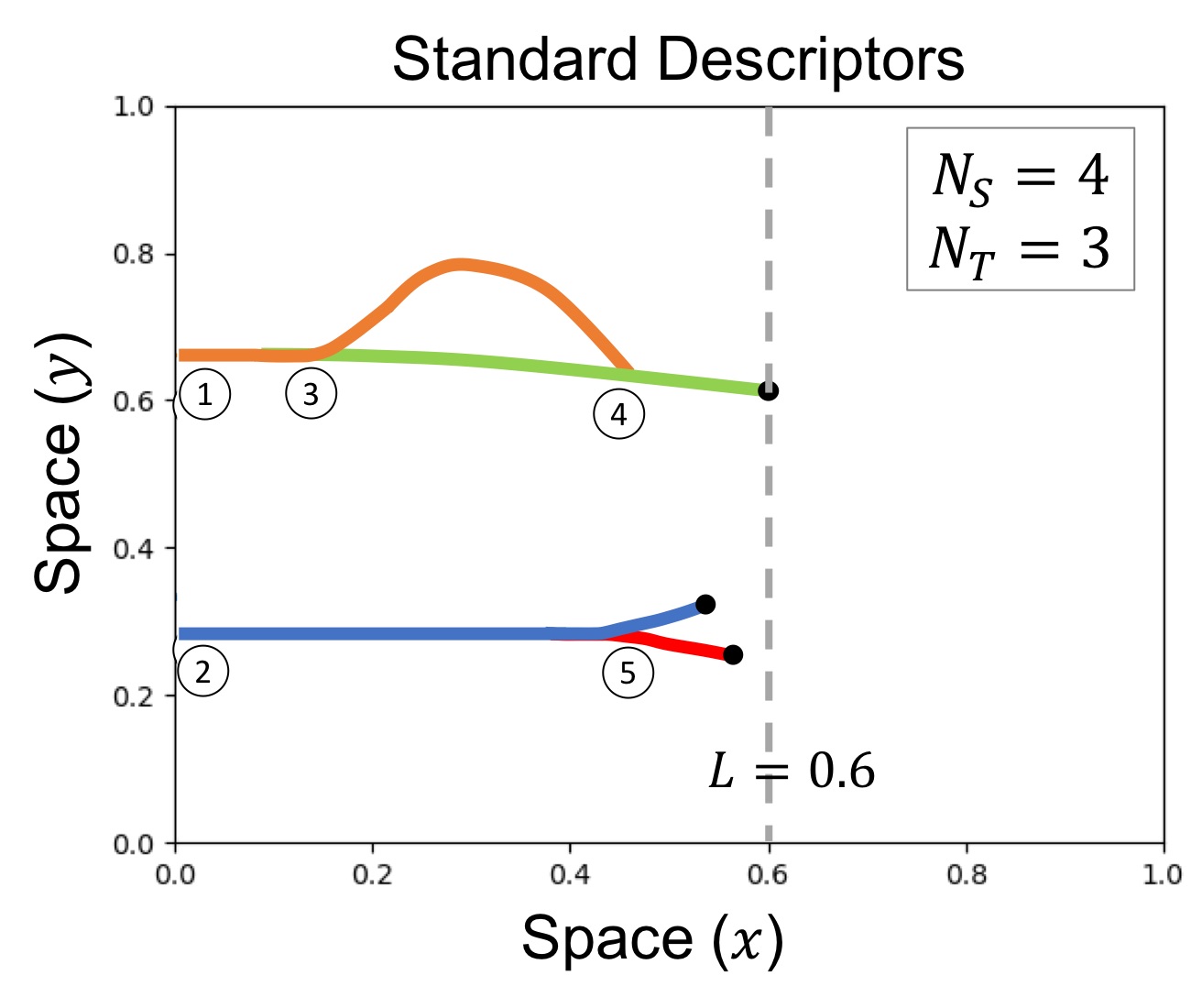}
    \caption{\textbf{Standard descriptor overview.} Standard descriptors are computed for each model simulation include the number of vessel segments ($N_S$, depicted by each colored solid curve), the number of endothelial tip cells ($N_T$, depicted by black dots), and the final length of the vasculature ($L$, depicted by the vertical grayed dashed line). This schematic example is initialized with 2 vessel segments and 2 tip cells at locations (1) and (2). A branching event at location (3) increases the value of both $N_S$ and $N_T$ by one. An anastomosis event at location (4) decreases $N_T$ by one. The branching event at location (5) increases both $N_S$ and $N_T$ by one.}
%    In practice, we count the number of vessel segments and tip cells at each simulation time step.}
    \label{fig:fig3}
\end{figure}

Morphologically-motivated standard descriptors are used to summarize physical characteristics of each model simulation. \new{Commonly-used descriptors include inter-vessel spacing, the number of branch points, the total area (volume) spanned by the network for 2D (3D) simulations, the shape of such loops in 2D (voids in 3D), and tortuosity ~\cite{baish_scaling_2011,folarin_three-dimensional_2010,kannan_functional_2018,konerding_evidence_1999,konerding_3d_2001, perfahl_3d_2017}.} 
\old{The standard descriptors that we} \new{We choose to use the following standard descriptors due to their prevalence in the literature} \old{use}: the number of vessel segments ($N_S$), the number of endothelial tip cells ($N_T$), and the final length of the longest vessel ($L$) \cite{balding_mathematical_1985,folarin_three-dimensional_2010,konerding_evidence_1999,konerding_3d_2001}. 
\new{In Fig~\ref{fig:fig3} we use idealised networks to indicate how the standard descriptors are defined. }
\old{Fig~\ref{fig:fig3} illustrates the standard descriptors at a snapshot of a schematic model simulation.} \new{As described in S1 Appendix, $N_S$ and $N_T$ both increase by one with each branching event, and  $N_T$ decreases by one with each anastomosis event.} We record $N_S$ and $N_T$ at 50 time steps to create \new{the} descriptor vectors. To facilitate downstream clustering analysis, we require vectors of the same length for all simulations. To ensure the dynamic descriptor vectors for each model simulation are of length 50, we record $N_S(\textbf{t})$ and $N_T(\textbf{t})$ at the 50 time steps $\bm{t}=\{t_i\}_{i=1}^{50}$, where $t_i = (i-1)\Delta t$, $\Delta t=t_\text{end}/49$, and $t_\text{end}$ denotes the duration of a given simulation. 
\new{We remark that the requirement for having vectors of the same length for all simulations might not be feasible when dealing with experimental data, which could include heterogeneities in sample size and time point locations. 
We note further that this requirement only applies to the standard descriptors; for the topological filtrations considered in future sections, only the final segmented vessel image is needed.} 
For each simulation, we also compute $L(t_\text{end})$, the $x$-coordinate of the vessel segment location which has the greatest horizontal distance from the $y$-axis. Thus, the standard descriptor vectors we use for subsequent clustering analysis are $N_S(\textbf{t}), N_T(\textbf{t})$ and $L(t_\text{end})$.

\subsection*{Quantifying vessel shape using topological data analysis}\label{subsec:PH}

The most prominent method from TDA is persistent homology (PH) \cite{carlsson_topology_2009,Edelsbrunner2010,Otter2017}. PH is rooted in the mathematical concept of homology, which captures the characteristics of shapes. To compute (persistent) homology from data, one needs to first construct simplicial complexes, which can be thought of as collections of generalized triangles. From the constructed simplicial complexes, one can quantify and visualize the connected components (dimension 0) and loops (dimension 1) of a dataset at different spatial scales. \new{While we compute the $N_T$ and $N_S$ standard decriptor vectors at 50 time points for each model simulation (see Section \textbf{Morphologically-motivated standard descriptors and descriptor vectors}), we compute topological descriptor vectors from the final output binary image.}

\subsubsection*{Simplicial complexes and homology} 
\begin{figure}
    \centering
    \includegraphics[width=0.99\textwidth]{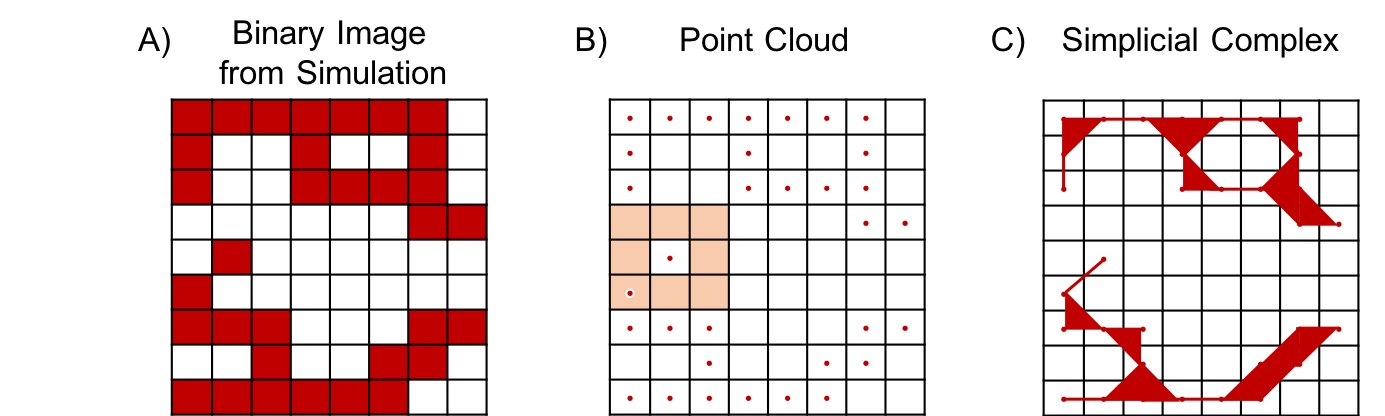}
    \caption{\textbf{Converting a binary image into a simplicial complex.} A) Schematic binary image. B) Point cloud for all nonzero pixels in the binary image. The Moore neighborhood of one nonzero pixel is highlighted in salmon.  C) A simplicial complex constructed from the point cloud in panel B. 
 }
    \label{fig:fig4}
\end{figure}

We construct simplicial complexes from data points derived from binary images of the completed simulations of the Anderson-Chaplain model (see``Synthetic data" in Step 1 of Fig~\ref{fig:fig2}, where red pixels denote presence of vasculature). Normally, TDA of image data lends itself to the construction of cubical complexes, which are generalizations of cubes; however, we have binary rather than grayscale data, so we instead use information about the model generation and Moore neighborhoods as we describe here.
Each pixel that has a value of one is embedded in $\mathbb{R}^2$ at the centroid of the pixel as a vertex (or $0$-simplex), as demonstrated in Fig.~\ref{fig:fig4}. We term the resulting collection of points a \emph{point cloud}. We connect two points by an edge (or $1$-simplex) if they are within each others' Moore neighborhood (the Moore neighborhood for one pixel is shaded in Fig.~\ref{fig:fig4}B). If three points are pairwise connected by an edge, then we connect them with a filled-in triangle (or $2$-simplex). The union of 0-,1-, and 2-simplices form a \emph{simplicial complex}, as shown in Fig.~\ref{fig:fig4}C.

 To compute topological invariants, such as connected components (dimension 0) and loops (dimension 1), we use homology. To obtain homology from a simplicial complex, $K$, we construct vector spaces whose bases are the $0$-simplices, $1$-simplices, and $2$-simplices, respectively, of $K$.  There is a linear map between $2$-simplices and $1$-simplices, called the boundary map $\partial_2$, which sends triangles to their edges. Similarly, the boundary map $\partial_1$ sends edges to their points and $\partial_0$ sends points to 0. The action of the boundary map $\partial_1$ on the simplices is stored in a binary matrix where the entry $a_{i,j}$ indicates whether the $i$-th $0$-simplex forms part of the boundary of the $j$-th $1$-simplex. If so, then $a_{i,j} = 1$; otherwise, $a_{i,j}=0$. One can compute the kernel $\text{Ker}(\cdot)$ and image $\text{Im}(\cdot)$ of the boundary maps to obtain the vector spaces $H_0(K) = \text{Ker}(\partial_{0}) / \text{Im}(\partial_{1})$ and
$H_1(K) = \text{Ker}(\partial_{1}) / \text{Im}(\partial_{2})$. These vector spaces are also referred to as homology groups and their dimensions define topological invariants called the \emph{Betti numbers} of $K$, $\beta_0$ and $\beta_1$, which quantify the numbers of connected components and loops, respectively.

\subsubsection*{Filtrations for simulated vasculature} 

There are different ways to study vascular data at multiple scales~\cite{bendich_persistent_2016,garside_topological_2019,Byrne2019,Stolz2020}. We encode the multiple spatial scales of the data using a \emph{filtration}, which is a sequence of embedded simplicial complexes $K_0\subseteq K_1 \subseteq \dots \subseteq K_\text{end}$ built from the data. \new{A challenge for researchers is to determine informative filtrations for specific applications \cite{stolz-pretzer_global_2019}, such as blood vessel development in our study.} Here, the input data is the binary image from the final timepoint of the Anderson-Chaplain model, which we call $\bm{N}$ (See Eq~(13) in S1 Appendix). We construct sequences of binary images that correspond to different filtered simplicial complexes: a sweeping plane filtration \cite{bendich_persistent_2016} and a flooding filtration \cite{garside_topological_2019}.
We remark that both filtrations can be considered sublevelset filtrations corresponding to a height function $h: X \rightarrow \mathbb{R}$ on some simplicial complex $X$ (or just on the vertices/pixels and considering  the simplicial subcomplex spanned by them).

\begin{figure}[ht!]
    \centering
    \includegraphics[width=0.99\textwidth]{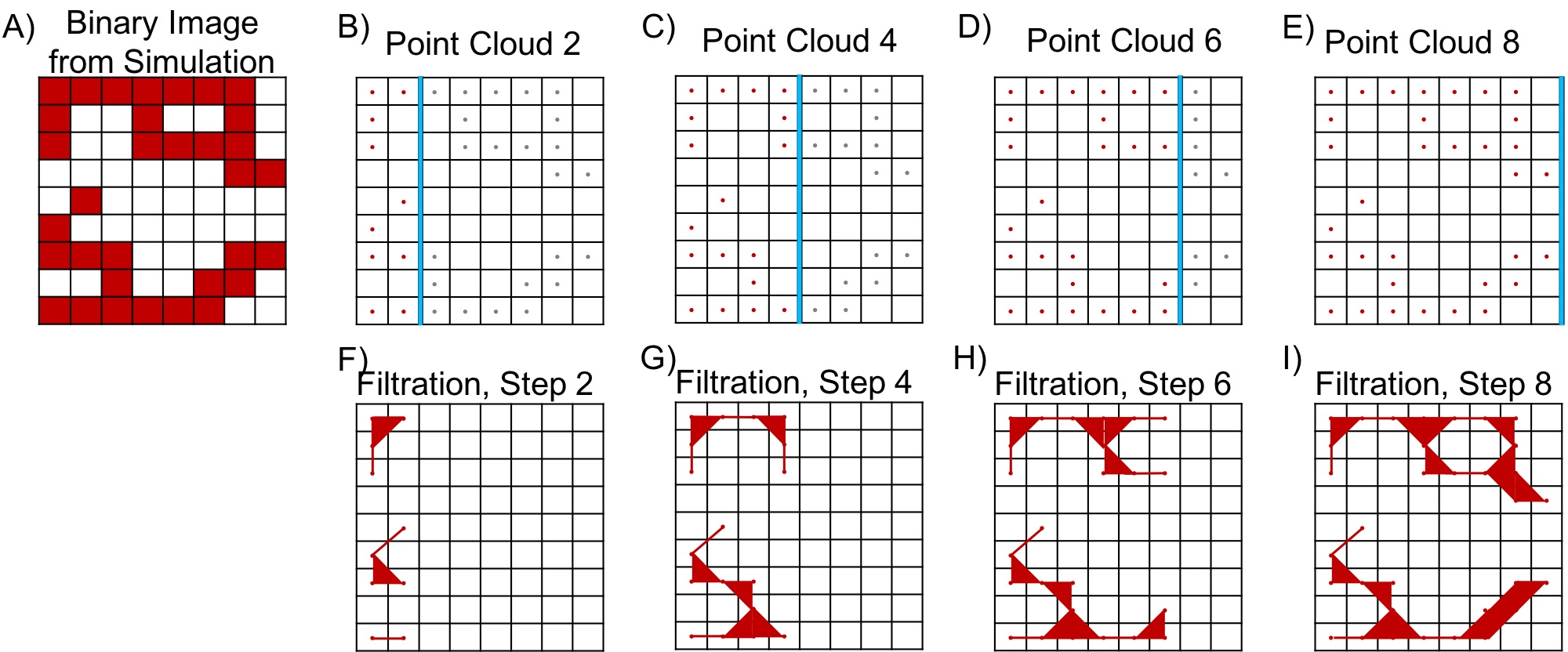}
    \caption{\textbf{The LTR sweeping plane filtration for a binary image.} A) Sample binary image from a simulation of the Anderson-Chaplain model. B-E) Point clouds used for each iteration of the sweeping plane approach. On each step, only pixels located to the left of the plane (denoted here with a blue line) are included; gray pixels are ignored. F-I) Filtration steps constructed from the point clouds in panels B-E.}
    \label{fig:fig5}
\end{figure}

\textbf{Sweeping plane filtration. }%Here we describe the sweeping plane approach
In the sweeping plane filtration, we move a line across the binary image $\bm{N}$ and include pixels in the filtration at discrete steps once a pixel is crossed by the line. On every filtration step, we move the line by a fixed number of pixels in a chosen direction. In Fig \ref{fig:fig5}, we illustrate the method when a vertical line moves from left to right (LTR). From the corresponding sequence of binary images, we construct a filtered simplicial complex $K^\text{LTR}=\{K_1,K_2,\dots,K_\text{end}\}$.  Similarly, we construct the filtered simplicial complexes $K^\text{RTL}$ when the vertical line moves from right to left (RTL), $K^\text{TTB}$ when a horizontal lines moves from top to bottom (TTB), and $K^\text{BTT}$ when a horizontal line moves from bottom to top (BTT). 

\begin{figure}[ht!]
    \centering
    \includegraphics[width=0.99\textwidth]{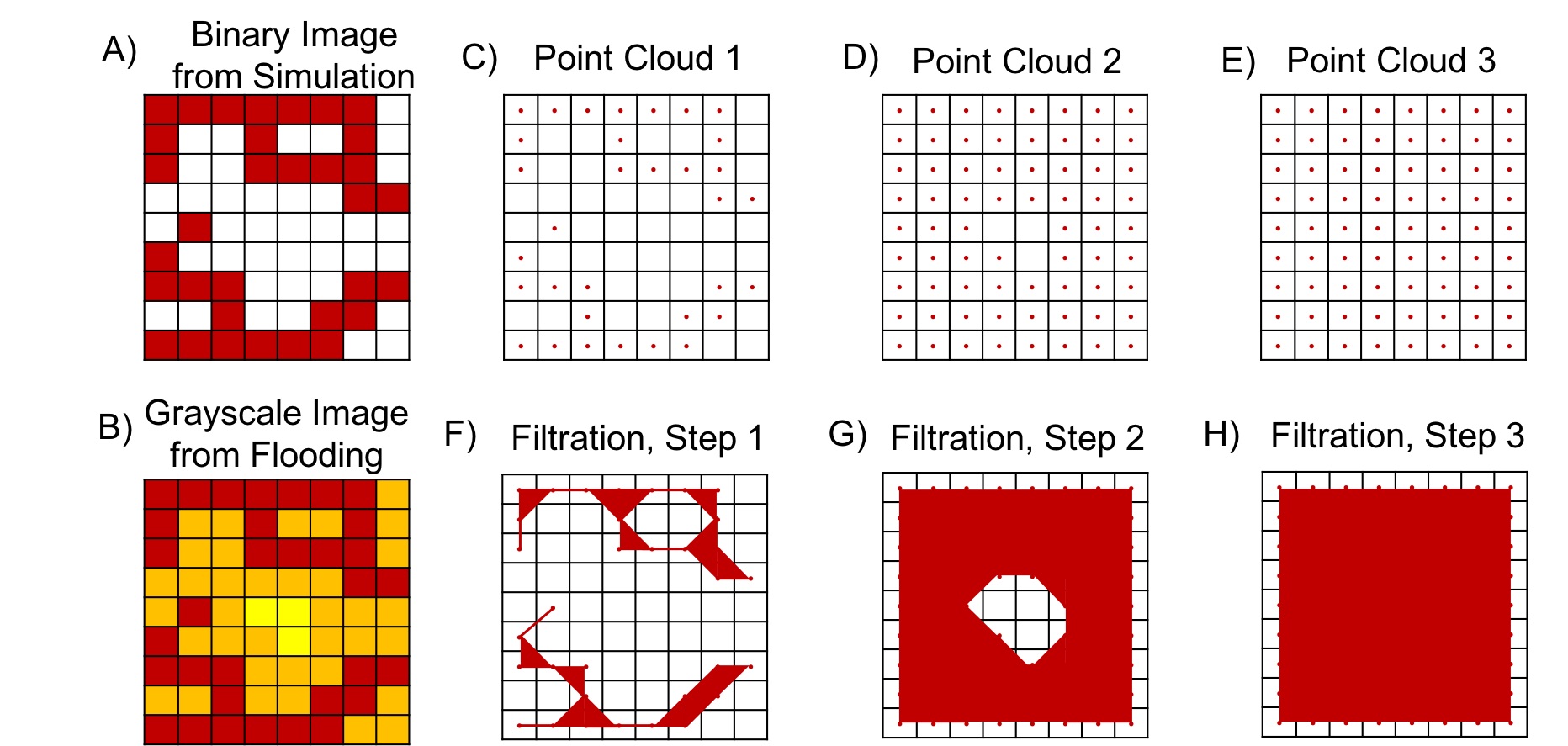}
    \caption{\textbf{The flooding filtration for a binary image.} A) Schematic binary image. B) Grayscale image of the original binary image. Red pixels are nonzero in the initial image, orange pixels become nonzero after one round of flooding, and yellow pixels become nonzero after two rounds of flooding. C-E) We performed three filtrations by dilating the image twice. During each filtration, the eight neighboring pixels of all nonzero pixels (ie pixels not marked as white) become nonzero. Red pixels are nonzero in the original image, orange pixels become nonzero on the second filtration (i.e., after one step of flooding), and yellow pixels become nonzero on the third filtration (i.e., after two steps of flooding). F-H) Simplicial complexes associated with the filtrations presented in panels C-E. }
    \label{fig:fig6}
\end{figure}

\textbf{Flooding filtration. }
Our second filtration is the flooding filtration. From the binary image, $\bm{N}$, associated with a model simulation, we construct a sequence of binary images in the following way. On the first step, we input the \old{final} output binary image from a model simulation. \new{In the binary image, pixels with value one denote the presence of an endothelial cell and pixels with value zero denote the absence of endothelial cells. To create the second binary image, we consider all pixels of value one and assign all pixels in their Moore neighborhood} (as shown in Fig~\ref{fig:fig4}B) to value one. We \new{repeat this process} \old{iterate} until \new{a binary image with only pixels of value one is created} \old{all pixels are set equal to one}. In Fig \ref{fig:fig6}B, for example, the red pixels denote the nonzero pixels from the initial binary image, the orange pixels denote the nonzero pixels added on the first round of flooding, and the yellow pixels denote the nonzero pixels added on the second round of flooding. Each ``round" is called a step. From the corresponding sequence of binary images, we construct a filtered simplicial complex $K^\text{flood}=\{K_1,K_2,\dots,K_\text{end}\}$. \new{We chose $K_\text{end}=25$ so that the Betti curves for all simulations attain $\beta_0=1$ and $\beta_1=0$ at the final filtration step, thus ensuring that the maximum amount of topological information is encoded in the Betti curves.}

\subsubsection*{Persistent homology (PH)} 
PH is an algorithm that takes in data via a filtration and outputs a quantification of topological features such as connected components (dimension 0) and loops (dimension 1) across the filtration. The simplicial complexes are indexed by the scale parameter of the filtration. In our case, the scale parameter corresponds to the spatial location in the sweeping plane filtration or steps of flooding in the flooding filtration. \new{Note that we compute these filtrations on simulated blood vessel network images at the final time step of a model simulation.} The inclusion of a simplicial complex $K_i\subseteq K_j$ for $i\leq j$ gives a relationship between the corresponding homology groups $H_p(K_i)$ and $H_p(K_j)$ for $p = 0,1$. This relationship enables us to track topological features such as loops along the simplicial complexes in the filtration. Intuitively, a topological feature is born in filtration step $b$ when it is first computed as part of the homology group $H_p(K_b)$ and dies in filtration step $d$ when that feature no longer exists in $H_p(K_d)$, \emph{i.e.}, when a connected component merges with another component or when a loop is covered by 2-simplices. The output from PH is a multiset of intervals $[b,d)$ which quantifies the persistence of topological features. Each topological feature is said to persist for the scale $d-b$ in the filtration. Our topological descriptors are the connected components (i.e., 0-dimensional) and loops (i.e., 1-dimensional features) obtained by computing persistent homology of the sweeping plane and flooding filtrations. We use the superscript notation $K^v$ to denote the filtered simplicial complex $K$ in direction $v=\{\text{LTR, RTL, TTB, BTT, flood}\}$.

\subsubsection*{Topological descriptor vectors from PH}
% Here we go from birth-death pairs to Betti-curves and persistence images
We use PH to generate different output vectors. 

\begin{itemize}
\item \textbf{Betti curves.} 
Betti curves~\cite{Giusti2015} show the Betti numbers on each filtration step. Let $\beta_i(K^v)$ denote the Betti curves of filtered simplicial complex $K$ in direction $v$ for dimensions $i=\{0,1\}$.

\item \textbf{Persistence images.} 
A persistence image~\cite{adams_persistence_2017} uses as input the birth-death pairs given by PH and converts the set of (birth $b$, persistence $(d-b)$) coordinates into a vector, a format which is suitable for machine learning and other classification tasks. \new{Following the standard definition of a persistence image, } \old{we begin by blurring} the coordinates $(b,d-b)$ \old{In a first step the coordinates $(b,d-b)$} are blurred by a Gaussian, with standard deviation $\sigma$, that is centered about each birth-persistence point,
\new{which accounts for uncertainty~\cite{adams_persistence_2017}.}
We take a weighted sum of the Gaussians and place a grid on this surface to create a vector. We compute two alternative weighting strategies: all ones (in which case all features are equally weighted) and a persistence weight of ($d-b$) (in which more persistent features are given larger weights). Let
%use the superscript notations 
PIO$_i(K^v)$ and PIR$_i(K^v)$ denote the corresponding persistence images computed from $K^v$ with equal and ramped weighting, respectively, 
%from the filtered simplicial complex $K$ in direction $v=\{\text{LTR, RTL, TTB, BTT, flood}\}$ 
for dimension $i=\{0,1\}$.
\end{itemize}

The topological descriptor vectors that we compute for each model simulation are $\beta_i(K^v)$, PIO$_i(K^v)$, and PIR$_i(K^v)$, where $i=\{0,1\}$ and $v=\{$LTR, RTL, TTB, BTT, flood $\}$. In total each simulation has $3 \times 2 \times 5 = 30$ topological descriptor vectors.

\subsection*{Simulation Clustering} 

We use an unsupervised algorithm, $k$-means clustering, to group the quantitative descriptor vectors of the synthetic data generated from simulations of the Anderson-Chaplain model. To cluster $n$ samples $x_1,x_2,\dots,x_n\in\mathbb{R}^d$ into $k$ clusters, the $k$-means algorithm finds $k$ points in $\mathbb{R}^d$, and clusters each sample according to which of the $k$ means it is closest to, and then minimizes the within-cluster residual sum of squares distance \cite{lloyd_least_1982}. We cluster the standard descriptor vectors and the topological descriptor vectors (see steps 2A and 2B of the methods pipeline in Fig~\ref{fig:fig2}), respectively, to investigate whether descriptor vectors lead to biologically \old{meaningful}\new{interpretable} clusters.

We test the robustness of each clustering assignment by randomly splitting the descriptor vectors into training and testing sets, and use the testing set to evaluate out-of-sample (OOS) accuracy. \new{Specifically, 7 of the 10 descriptor vectors from each of the 121 $(\rho,\chi)$ parameter combinations are randomly chosen with uniform sampling and placed into a training set (847 simulations); the remaining 3 descriptor vectors from each $(\rho,\chi)$ combination are placed into a testing set (363 simulations).} After performing unsupervised clustering with a $k$-means model on the training set, each $(\rho,\chi)$ combination is labeled by the majority of cluster assignments among its 7 training simulation descriptor vectors.  To evaluate each clustering model's OOS accuracy, the 3 test data samples for each $(\rho,\chi)$ parameter combination are given a ``ground-truth'' label equal to the cluster assignment for the 7 training simulations for that same $(\rho,\chi)$ parameter combination. A ``predicted'' label is calculated for each descriptor vector from the test data using the trained $k$-means model, and OOS accuracy is defined as the proportion of test simulations for which the ``predicted'' label matched the ``ground-truth'' label. For example, if the seven training simulations from the parameter combination $(\rho,\chi)=(0.2,0.2)$ are placed into clusters $\{1,1,2,2,1,1,1\}$, then the three testing simulations from $(\rho,\chi)=(0.2,0.2)$ will be labeled as being in cluster 1. If the test simulations are predicted to be from clusters $\{1,2,1\}$, then we compute an OOS accuracy of 66.67$\%$. We use all 363 testing simulations when computing OOS accuracy and \new{use the \texttt{KMeans} command from scikit-learn (version 0.22.1), which uses Lloyd's algorithm for Euclidean space, for training and prediction.}

\section*{Results}

We simulated the Anderson-Chaplain model for 121 different values of the
chemotaxis and haptotaxis coefficients ($ \rho, \chi$) and performed 10 model realizations for each parameter pair. We observed that simulations generated by different parameter pairs may appear visually distinct in their vessel growth patterns. To quantify and summarize the morphologies, we computed dynamic standard descriptor vectors as well as topological descriptor vectors for the set of 1,210 model simulations. We then performed unsupervised clustering using either standard or topological descriptor vectors to partition the $(\rho,\chi)$ parameter space. 

As we detail here, we found that the topological sweeping plane descriptor vectors offered a multi-scale analysis that led to biologically interpretable and robust clusterings of the $(\rho,\chi)$ parameter space, whereas the clusterings of the standard descriptor vectors were not biologically interpretable. The topological analysis succeeded in discriminating fine-grained vascular structures and the parameter pairs that generated them.

\subsection*{Haptotaxis and chemotaxis alter vessel morphology}

%\new{[Previous reviews already documenting how hapt and chemotaxis affect morphology \cite{oden_toward_2016,hadjicharalambous_tumour_2021}]}
%\new{[Previous studies to cite on mechanotaxis \cite{vilanova_computational_2018,vavourakis_validated_2017}]}

\begin{figure}[ht!]
    \centering
    \includegraphics[width=0.99\textwidth]{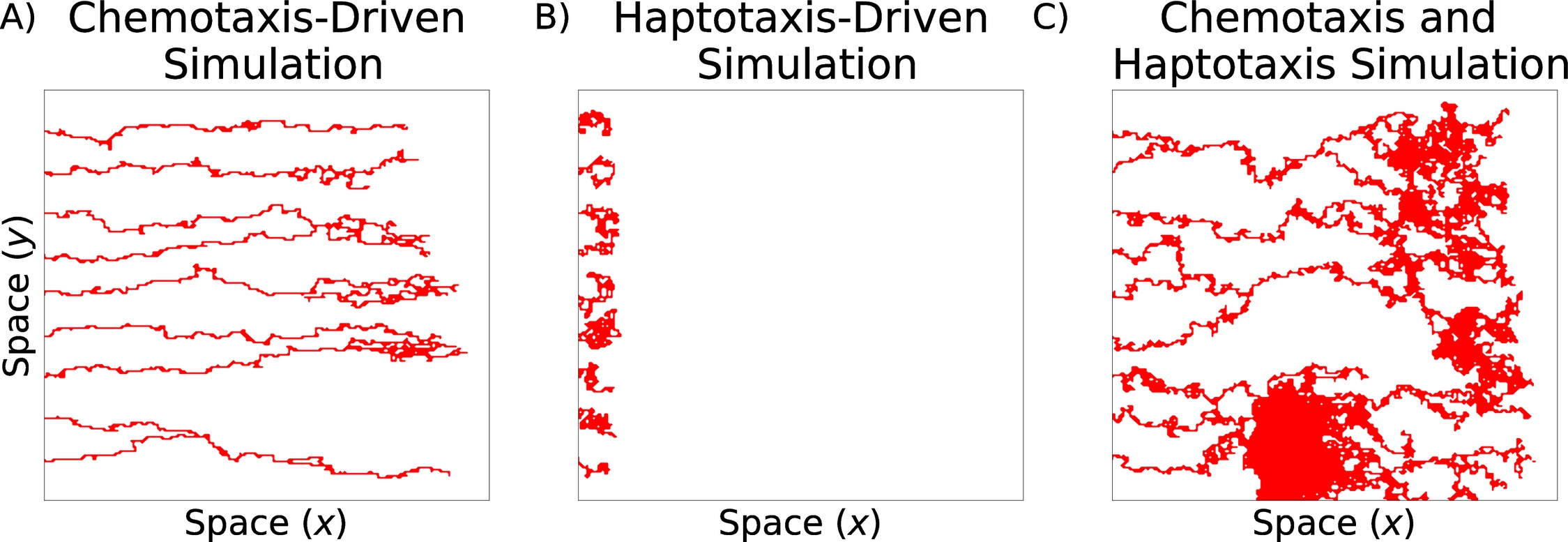}
    \caption{\textbf{Typical realizations of the Anderson-Chaplain model}. A)  Chemotaxis-driven tip cell movement ($\rho=0.00$, $\chi=0.50$), B) a haptotaxis-driven tip cell movemement ($\rho=0.50$, $\chi=0.00$), and C) tip cell movement driven by chemotaxis and haptotaxis ($\rho=0.15,\chi=0.15$).}
    \label{fig:fig7}
\end{figure}

We investigated the influence of haptotaxis and chemotaxis by varying the parameters $\rho$ and $\chi$ \old{, respectively}. \new{We note that similar parameter sensitivity analyses have been performed previously \cite{anderson_continuous_1998,hadjicharalambous_tumour_2021}; we reproduce this analysis as the basis for our data analysis.} \old{for consistency in our study.} Recall that in model simulations, the vessels emerged from the left-most boundary and were recruited towards a tumor located on the right boundary. 
When chemotaxis alone drives endothelial tip cell movement ($\rho=0.00$,$\chi = 0.50$), the blood vessel segments are long and thin as the tip cells migrate directly towards the tumor and produce individual vessel segments with few anastomosis events (Fig~\ref{fig:fig7}A). 
When haptotaxis drives tip cell movement ($\rho = 0.50$,$\chi = 0.00$),
the vessels fail to reach the tumor (Fig~\ref{fig:fig7}B). When both haptotaxis and chemotaxis are active  ($\rho=\chi=0.15$), the network spans a large portion of the spatial domain and is highly connected (Fig~\ref{fig:fig7}C). 

\subsection*{Standard and topological descriptors measure distinct aspects of vascular architectures}

\begin{figure}
    \centering
    \includegraphics[width=.92\textwidth]{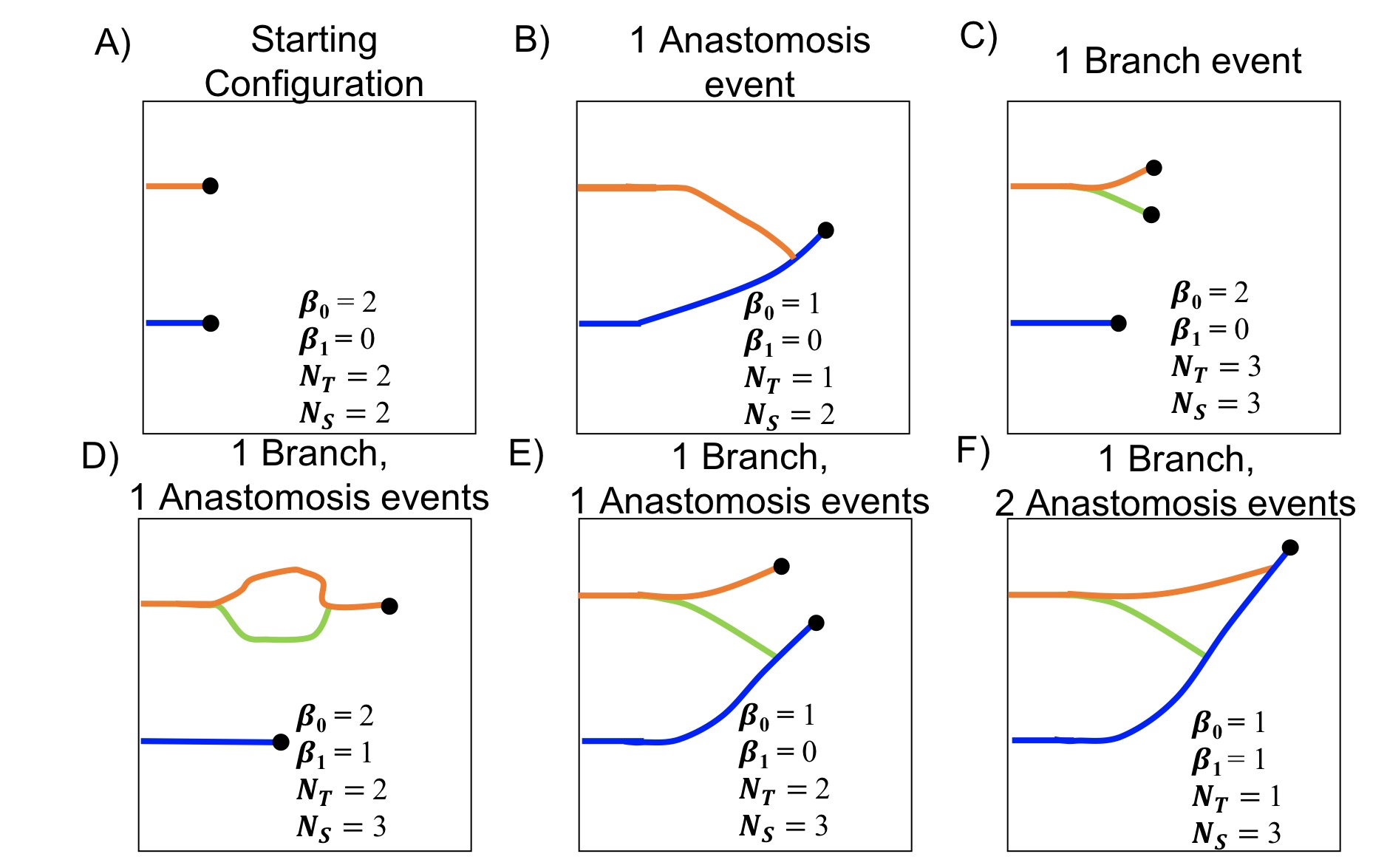}
    \caption{\textbf{Standard and topological descriptors describe distinct aspects of vessel morphology}. For clarity, each vessel segment is colored differently, and black dots represent active tip cells. A) The initial vessel configuration. B) A schematic showing how the vascular architecture changes after an anastomosis event.
    C) A schematic showing how the vascular architecture changes after a branching event. D-E) Schematics showing two ways in which vasculature can change after one anastomosis event and one branching event. F) A schematic showing how the vessels can change after one branching and two anastomosis events. (Key: $\beta_0$, number of connected components; $\beta_1$, number of loops; $N_T$, number of tip cells; $N_S$, number of vessel segments).}
    \label{fig:fig8}
\end{figure}

Rather than performing visual inspection of all model simulations, we computed and compared quantitative vessel descriptor vectors using data science approaches as described in the \textbf{Materials and Methods}. 
We demonstrate here how the biological and topological descriptors are distinct from each other. In Fig~\ref{fig:fig8}, we observe how the number of tip cells ($N_T$), vessel segments ($N_S$), connected components ($\beta_0$), and loops ($\beta_1$) associated with an evolving vasculature change following different events. After each anastomosis event, $N_T$ decreases by one (Fig~\ref{fig:fig8}B).  After a branching event, both $N_T$ and $N_S$ increase by one (Fig~\ref{fig:fig8}C). We conclude that $N_T$ is unchanged and $N_S$ increases by one after one branching event and one anastomosis event have occurred. By contrast, the values of the topological descriptors before and after these events are not necessarily identical, as depicted in Figs \ref{fig:fig8}D-E. When an anastomosis event involves vessel segments that were previously connected, the number of loops, $\beta_1$, increases by 1 (compare Figs \ref{fig:fig8}A and \ref{fig:fig8}D); when the vessel segments were not previously connected, $\beta_0$ decreases by 1 (compare Figs \ref{fig:fig8}A and E).

Let $N_A(r,s)$ and $N_B(r,s)$ denote the numbers of anastomosis and branching events, respectively, that occur between times $t=r$ and $t=s$. We computed $N_T(\bm{t}), N_S(\bm{t})$ vectors of the standard descriptors of evolving blood vessel simulations as follows:
\begin{align*}
    N_T( t_1 ) &= N_S(t_1) = 9, \\
    N_T( t_i ) &= N_T(t_{i-1}) + N_B(t_{i-1},t_i) - N_A(t_{i-1},t_i),\ i = 2,...,50, \\
    N_S(t_i) &= N_S(t_i) + N_B(t_{i-1},t_i),\ i = 2,...,50.
\end{align*}
The values of $N_T(t_1)$ and $N_S(t_1)$ were determined from the models' initial conditions. 

We first constructed the filtrations $K^v$ where $v=\{\text{LTR, RTL, TTB, BTT, flood}\}$ of each simulation's final binary image. Then we computed connected components and loops, which we call topological descriptors of $K^v$, using PH. The topological descriptor vectors, Betti curves and persistence images, were then constructed for these five filtrations.

\subsection*{Quantification of blood vessel architecture data}

\begin{figure}
    \centering
    \includegraphics[width=\textwidth]{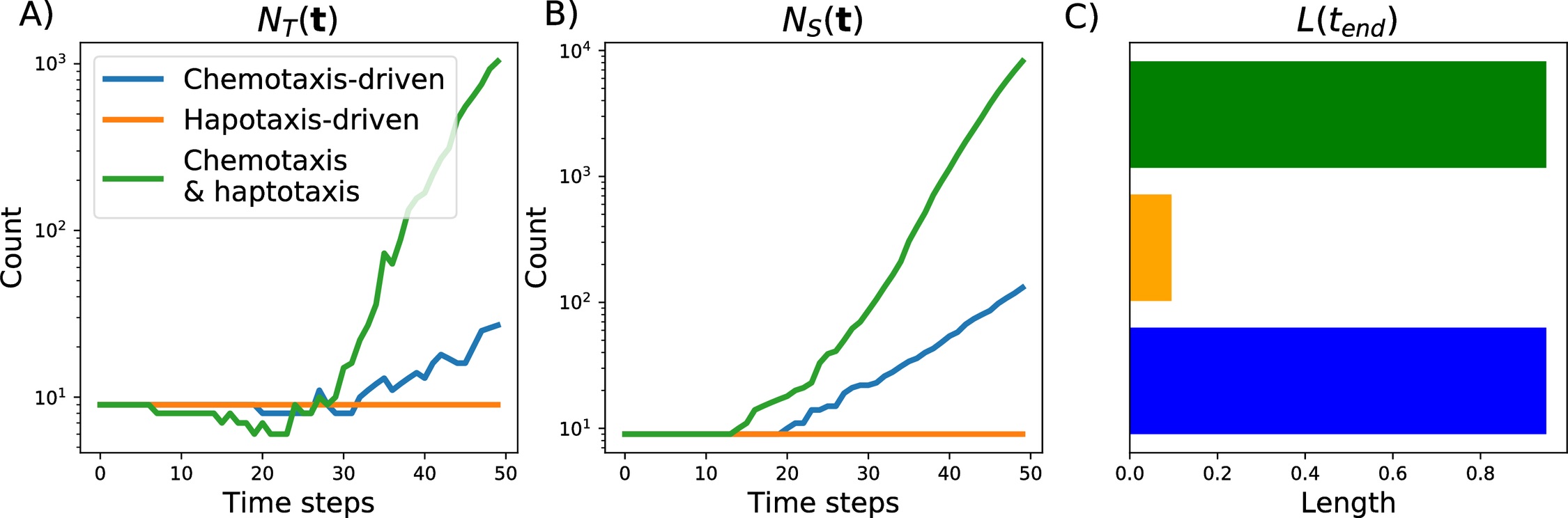}
    \caption{\textbf{The standard descriptors used to summarize model simulations.} We show how the following descriptors change over 50 time steps for the three simulations presented in Fig~\ref{fig:fig7}: A) the number of tip cells, $N_T(\textbf{t})$, B) the number of vessel segments, $N_S(\textbf{t})$, and C) the final vessel length, $L(t_{end})$.}
    \label{fig:fig9}
\end{figure}

Fig~\ref{fig:fig9} shows the standard descriptor vectors for the three simulations presented in Fig~\ref{fig:fig7}. For the haptotaxis-driven simulation growth in the horizontal direction halts (the total length remains at 0.095 units) and the numbers of tip cells and vessel segments stay constant (at or near the initial condition of 9 vessels). By contrast, the number of tip cells and vessel segments increase over time for the chemotaxis-driven simulation, with a more marked increase when both chemotaxis and haptotaxis are active. For both chemotaxis-active simulations, the vessels extend to the maximum horizontal distance.

\begin{figure}[ht!]
    \centering
    \includegraphics[width=\textwidth]{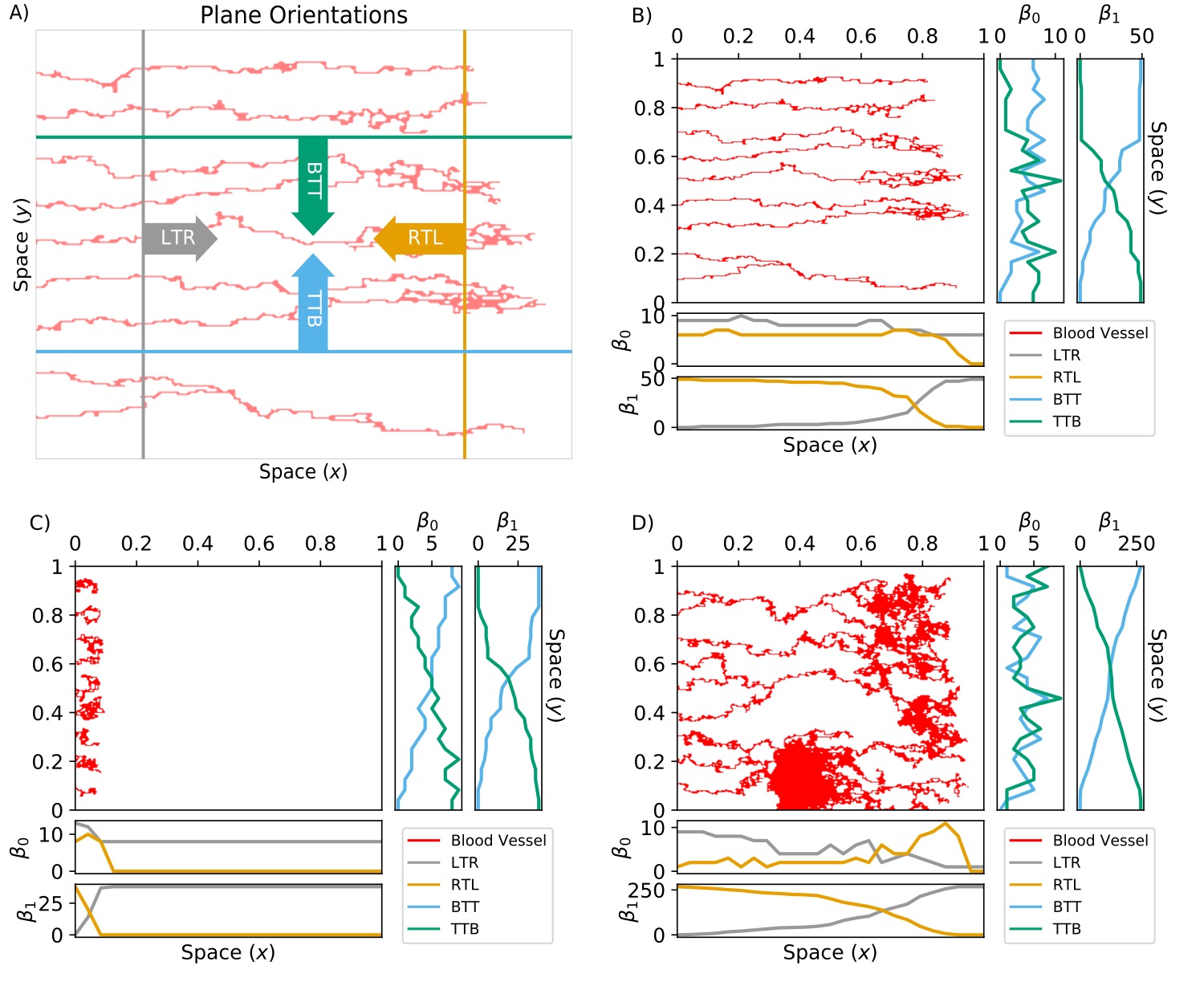}
    \caption{TDA sweeping plane descriptor vectors. A) The sweeping plane directions are left-to-right (LTR, gray); right-to-left (RTL, orange); bottom-to-top (BTT, cyan); and top-to-bottom (TTB, green). For these four \emph{v}=\emph{i}T\emph{j} filtrations, we started with points on the \emph{i}th boundary and included more points as we stepped towards the \emph{j}th boundary. Repeating this process produced the spatial filtration $K^v=K_0,K_1\ldots K_{\text{end}}$ for $v=\{\text{LTR, RTL, TTB, BTT}\}$ . B-D) We formed the Betti curves $\beta_0(K^v)$ and $\beta_1(K^v)$ by computing the Betti numbers along each step of $K^v$ for the three blood vessel simulations presented in Fig \ref{fig:fig7}; B) the chemotaxis-driven simulation, C) the haptotaxis-driven simulation and D) chemotaxis \& haptotaxis simulation}.
    \label{fig:fig10}
\end{figure}

\begin{figure}
    \centering
    \includegraphics[width=\textwidth]{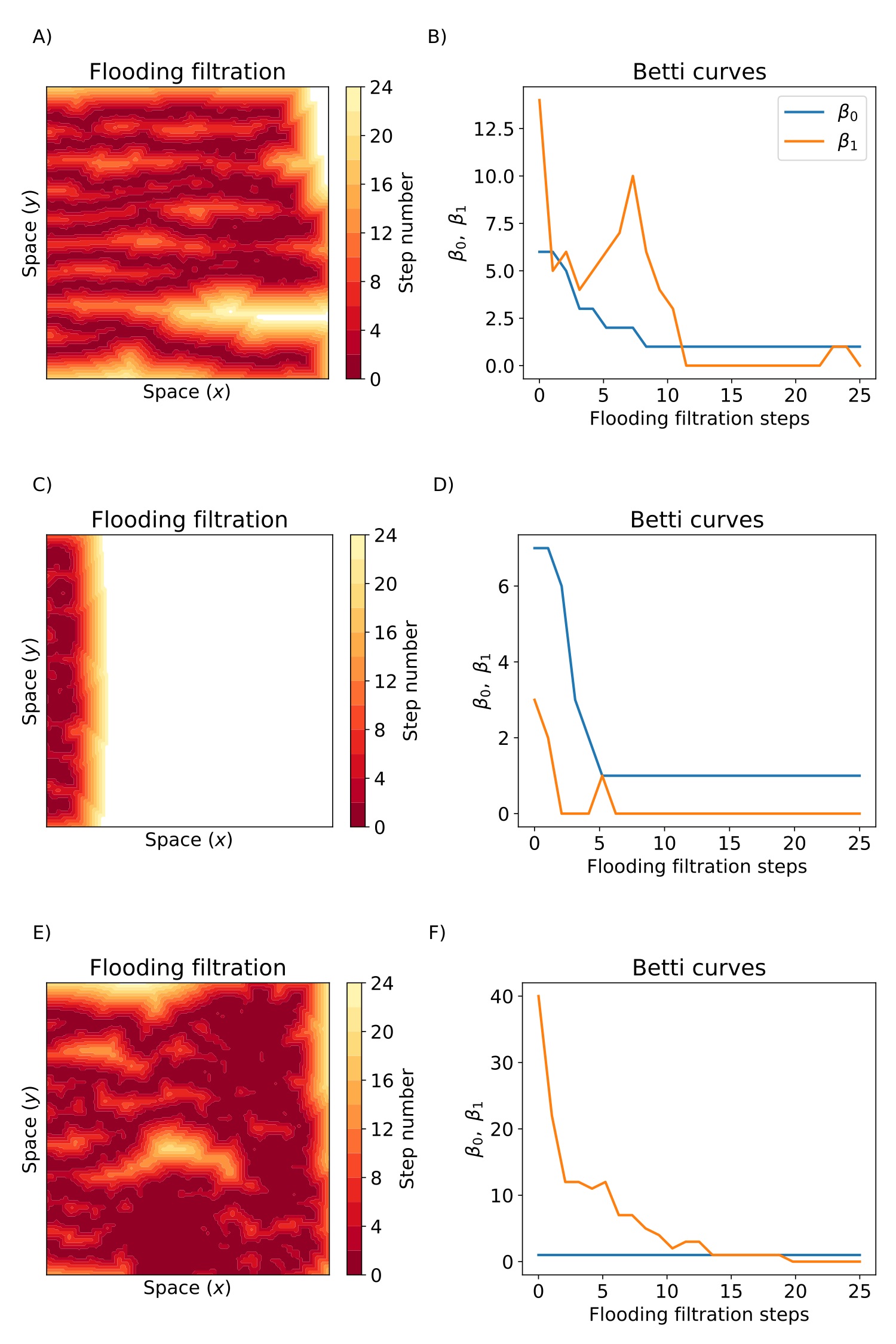}
    \caption{\textbf{The flooding filtration.} A,C,E) Illustrate the flooding filtrations for the three blood vessel simulations from Fig \ref{fig:fig7}. Dark red pixels were points included in the first steps of the filtration and yellow pixels were the points included in later filtration steps. White pixels were not included after 24 steps of the flooding filtration.   B,D,F) The Betti Curves ($\beta_0(K^{\text{flood}})$ and $\beta_1(K^{\text{flood}}$)  for the chemotaxis-driven, haptotaxis-driven, and chemotaxis \& haptotaxis simulations respectively.}
    \label{fig:fig11}
\end{figure}

We next illustrate \old{the topological analysis for} \new{how we use the plane sweeping and flooding filtrations to analyse} the chemotaxis-driven, haptotaxis-driven, and chemotaxis \& haptotaxis-driven simulations (from Fig~{\ref{fig:fig7}}). We  computed Betti curves, $\beta_0(K^v)$ and $\beta_1(K^v)$, of the simulated data with sweeping plane filtrations for $v=\{\text{LTR, RTL, TTB, BTT}\}$ (see Fig~\ref{fig:fig10}) and a flooding filtration $v=\{\text{flood}\}$ (see Fig~\ref{fig:fig11}). 
\new{These filtrations are constructed to provide complementary but not directly comparable information; roughly, the sweeping plane gives an indication of the network in the (x-y) plane whereas the flooding filtration also focusses on vessel density (see earlier description of filtration construction).}
%\new{Since the plane sweeping approach is filtered over the position of a plane while the flooding approach is filtered over the round of flooding steps, these biological descriptor vectors cannot be depicted on the same axis.}  
We could interpret the PH of the sweeping plane filtration direction
\old{. Since}\new{ as follows:} the blood vessels grow primarily from left to right,  \new{so the LTR and RTL filtrations}\old{the topological analysis} primarily identify dynamic changes in branching and anastomosis of vessels. The BTT and TTB filtrations are similar to each other, and reflect only horizontal changes, which could be interpreted as measuring bendiness or tortuosity.  Therefore, we only discuss the BTT results in this section.

For the chemotaxis-driven simulation (Figs~\ref{fig:fig10}B, \ref{fig:fig11}A, and \ref{fig:fig11}B), the $\beta_0(K^\text{LTR})$ descriptor vector slightly decreases with distance $x$ from the $y$-axis as vessel segments anastomose together. The $\beta_0(K^\text{RTL})$ and $\beta_1(K^\text{RTL})$ descriptor vectors decrease with $x$ at the end of each vessel segment. The magnitude of $\beta_0(K^{\text{BTT}})$ increases almost linearly with the $y$-spatial coordinate, and $\beta_1(K^{\text{BTT}})$ increases with $y$ until it plateaus for $y \geq 0.6$. The $\beta_0(K^{\text{flood}})$ descriptor vector decreases with each filtration step, and $\beta_1(K^{\text{flood}})$ begins with $\beta_1=14$ and achieves a local maximum of $\beta_1=10$ after seven filtration steps before reaching $\beta_1=0$ after 11 filtration steps. 

For the haptotaxis-driven simulation (Figs ~\ref{fig:fig10}C,  \ref{fig:fig11}C, and \ref{fig:fig11}D), all LTR and RTL Betti curves become zero for $x>0.1$ because each vessel segment has terminated by $x=0.1$. The $\beta_0(K^{\text{BTT}})$ and $\beta_1(K^{\text{BTT}})$ descriptor vectors increase steadily with $y$ as the horizontal plane moves past each individual vessel segment. The flooding descriptor vectors initially decrease with each flooding step and remain fixed at $\beta_0=1,\beta_1=0$ after six steps of flooding.

For the chemotaxis \& haptotaxis-driven simulation (Figs \ref{fig:fig10}D,  \ref{fig:fig11}E, \ref{fig:fig11}F), $\beta_0(K^\text{LTR})$ decreases with $x$ as vessel segments anastomose together, while $\beta_1(K^\text{LTR})$ increases with $x$ due to the formation of many loops. The $\beta_0(K^\text{RTL})$ descriptor vector increases with $x$ until it reaches its maximum value of $\beta_0=11$ at $x=0.85$ and then decreases to $\beta_0=0.$ The $\beta_1(K^\text{RTL})$ descriptor vector decreases with $x$. The $\beta_0(K^\text{BTT})$ descriptor vector periodically increases and decreases with $y$ as the horizontal plane moves past regions of high and low connectivity. The $\beta_1(K^\text{BTT})$ descriptor vector steadily increases with the $y$ coordinate. The $\beta_0(K^\text{flood})$ descriptor vector was equal to one for all steps of flooding, and $\beta_1(K^\text{flood})$ decreases with each round of flooding and reaches $\beta_1=0$ after 20 flooding steps.
 
\subsection*{Sweeping plane descriptor vectors cluster simulations by parameter values}
% Clustering results from different summaries

\begin{figure}
    \centering
    \includegraphics[width=.89\textwidth]{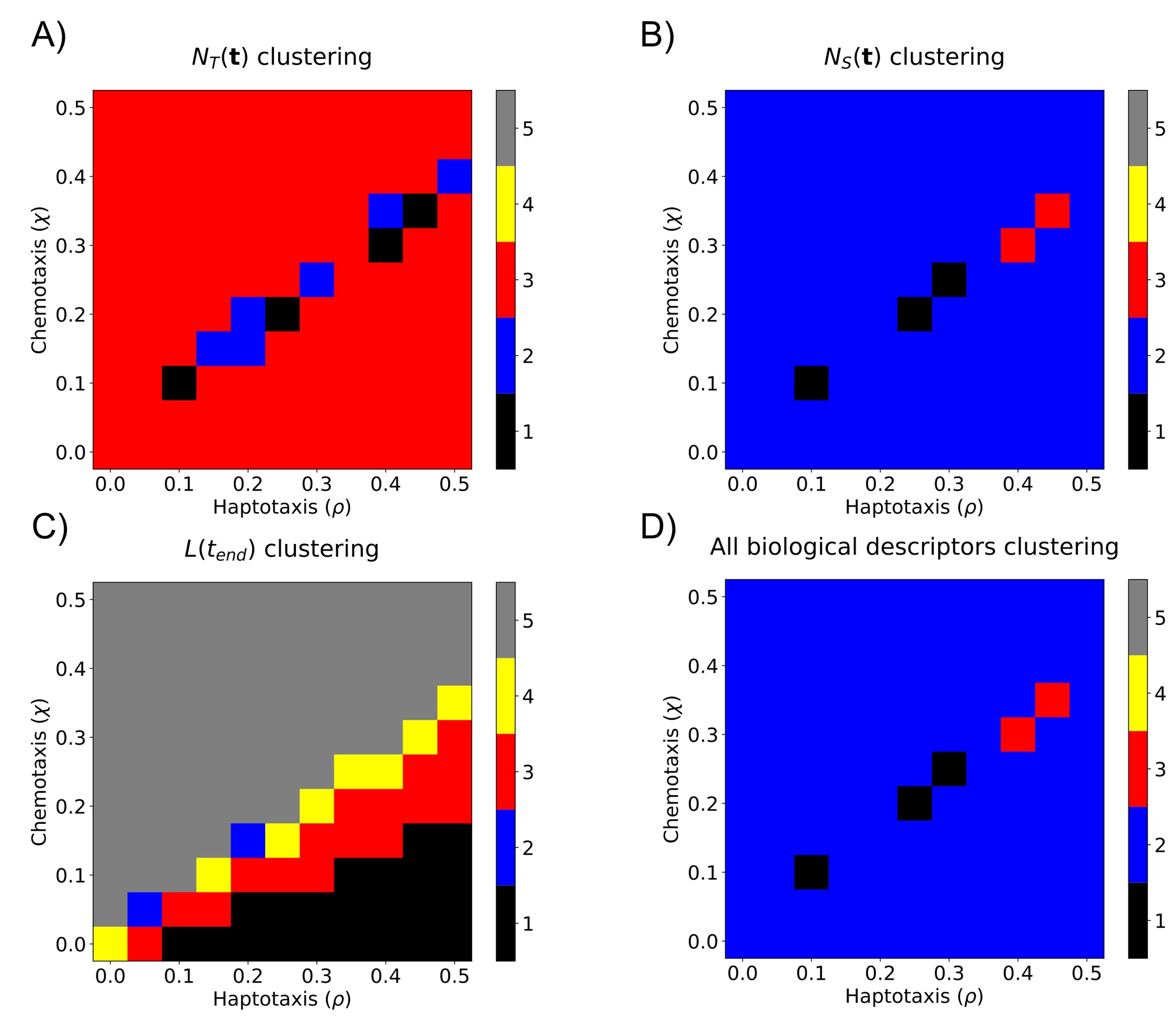}
    \caption{\new{\textbf{Partitioning $(\rho,\chi)$ parameter space into distinct regions using standard biological descriptor vectors.} We applied $k$-means clustering, with $k=5$, to standard biological descriptor vectors. The resulting partitions are presented for  A) $N_T(\bm{t})$, B) $N_S(\bm{t})$, C) $L(t_{end})$, and D) concatenation of all three descriptor vectors. The five clusters are ordered according to the mean $\chi$ value within the cluster.}
}
    \label{fig:fig12}
\end{figure}

We applied $k$-means clustering separately to the standard descriptor vectors and the topological descriptor vectors. None of the clusterings of the standard descriptor vectors were biologically \new{interpretable}\old{meaningful} in ($\rho$,$\chi$) parameter space (see \new{Fig \ref{fig:fig12} and}  S\ref{fig:SIfig1} Fig.) While the topological descriptor vectors from the sweeping plane filtration gave more biologically interpretable groupings for combined haptotaxis and chemotaxis parameter values, they were not robust for haptotaxis dominated or chemotaxis dominated  ($\rho$,$\chi$) parameter ranges (see S\ref{fig:SIfig2} Fig).  The flooding filtration did not give biologically \old{meaningful}\new{interpretable} groups for any parameter ranges (see \textbf{Clustering results} in S1 Appendix for all details). \new{We suggest that the LTR and RTL plane sweeping filtrations produce the most interpretable clusterings because sweeping the plane horizontally resembles the way in which endothelial tip cells migrate from left to right in the simulated data. }

\new{No single descriptor vector was able to robustly cluster simulations into biologically interpretable groups.} However, \emph{double} descriptor vectors created by concatenating two sweeping plane topological descriptor vectors\footnote{For example, the double descriptor vector ``PIO$_0(K^{\text{LTR}})$ \& PIO$_1(K^{\text{LTR}})$'' denoted the topological descriptor vector created by concatenating the PIO$_0(K^{\text{LTR}})$ and PIO$_1(K^{\text{LTR}})$ descriptor vectors.} produced clusterings in ($\rho$,$\chi$) parameter space that were biologically interpretable and robust. We note that doubles of standard descriptor vectors or topological flooding descriptor vectors were unable to produce biologically interpretable and robust clusterings in $(\rho,\chi)$ parameter space (see \textbf{Clustering results} in S1 Appendix). 

\begin{figure}
    \centering
    \includegraphics[width=0.89\textwidth]{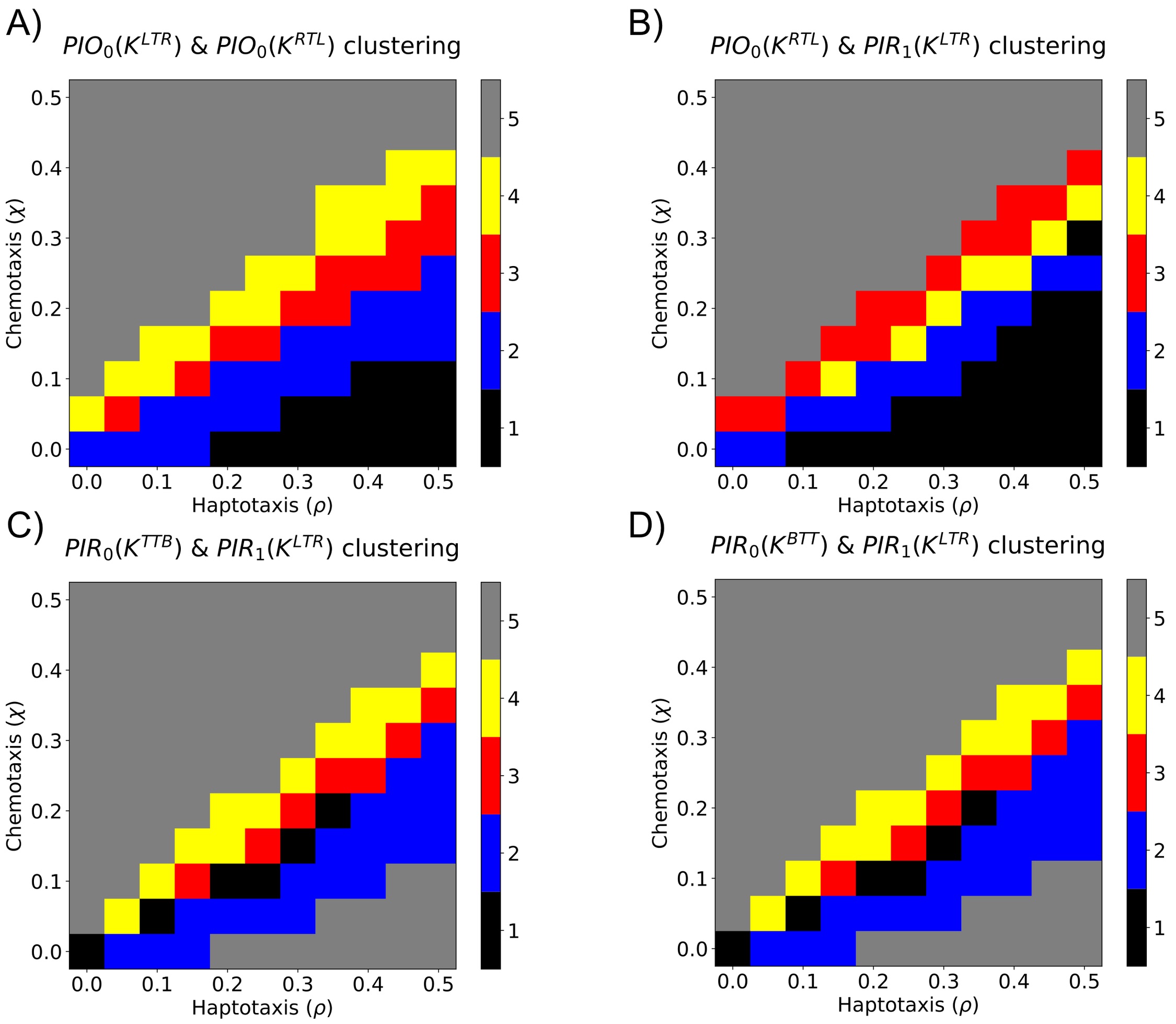}
    
    \caption{\textbf{Partitioning $(\rho,\chi)$ parameter space into distinct regions using sweeping plane double descriptor vectors.} We applied $k$-means clustering, with $k=5$, to sweeping plane double descriptor vectors. The resulting partitions are presented for the four descriptor vectors that have the highest OOS accuracy: A) $PIO_0(K^{LTR})$ \& $PIO_0(K^{RTL})$, B) $PIO_0(K^{RTL})$ \& $PIR_1(K^{LTR})$, C) $PIO_0(K^{TTB})$ \& $PIR_1(K^{LTR})$, and D) $PIO_0(K^{BTT})$ \& $PIR_1(K^{LTR})$. The five clusters are ordered according to the mean $\chi$ value within the cluster.}
    \label{fig:fig13}
\end{figure}

We computed a total of $2\times4\times3=24$ individual sweeping plane descriptor vectors for each model simulation: each descriptor vector counted connected components or loops ($\beta_0$, $\beta_1$), swept a plane across the domain in one of four directions (LTR, RTL, TTB, and BTT), and used one of the three PH summaries (BC, PIO, PIR). We computed all 
%$(24\text{ choose }2)=276$ 
double sweeping plane descriptor vectors and clustered all 276 combinations in $(\rho,\chi)$ parameter space. The 15 double descriptor vectors with the highest OOS accuracies are presented in Table~\ref{tab:tab1}; the four with the highest OOS accuracies are liested below:
\begin{enumerate}
    \item PIO$_0(K^{LTR})$ \& PIO$_0(K^{RTL})$ (93.1\% OOS accuracy),
    \item PIO$_0(K^{RTL})$ \& PIR$_1(K^{LTR})$ (92.6\% OOS accuracy),
    \item PIR$_0(K^{TTB})$ \& PIR$_1(K^{LTR})$ (91.5\% OOS accuracy), 
    \item PIR$_0(K^{BTT})$ \&  PIR$_1(K^{LTR})$ (91.5\% OOS accuracy).
\end{enumerate}
In Fig~\ref{fig:fig13}, we show how the data 
clustered in 
($\rho,\chi$) parameter space when the top four double sweeping plane descriptor vectors were used. The ``PIO$_0(K^{LTR})$ \& PIO$_0(K^{RTL})$" and  ``PIO$_0(K^{RTL})$ \& PIR$_1(K^{LTR})$" doubles produced five connected clustering regions and, thus, attained high OOS accuracy and biologically interpretable clusterings. In contrast, the clusterings generated by the following double descriptor vectors (``PIR$_0(K^{TTB})$ \& PIR$_1(K^{LTR})$" and ``PIR$_0(K^{BTT})$ \& PIR$_1(K^{LTR})$")  \old{was}\new{were} unable to distinguish between simulations with strong chemotaxis and strong haptotaxis. 
\new{We suggest that double descriptor vectors combining LTR and RTL descriptor vectors can robustly cluster simulations into biologically interpretable clusterings because the associated filtrations capture the ways in which vessel segments anastomose and branch, respectively. We performed principal components analysis to reduce the ``PIO$_0(K^{LTR})$ \& PIO$_0(K^{RTL})$" double descriptor vector to two principal components. The results presented in S\ref{fig:SIfig3} Fig show that the simulations from each group are also clustered in this two-dimensional space.}

\begin{table}
    \centering
    \def\arraystretch{1.5}
    \begin{tabular}{|c|c|c|}
\hline
Feature & In Sample Accuracy & Out of Sample Accuracy \\ 
\hline
PIO$_0(K^{LTR})$ \& PIO$_0(K^{RTL})$ & 94.0\% & 93.1\% \\
\hline
PIO$_0(K^{RTL})$ \& PIR$_1(K^{LTR})$ & 94.2\% & 92.6\% \\
\hline
PIR$_0(K^{TTB})$ \& PIR$_1(K^{LTR})$ & 92.3\% & 91.5\% \\
\hline
PIR$_0(K^{BTT})$ \& PIR$_1(K^{LTR})$ & 92.3\% & 91.5\% \\
\hline
PIR$_0(K^{LTR})$ \& PIR$_1(K^{LTR})$ & 92.3\% & 90.9\% \\
\hline
PIO$_0(K^{TTB})$ \& PIR$_1(K^{LTR})$ & 94.2\% & 90.4\% \\
\hline
PIO$_0(K^{BTT})$ \& PIR$_1(K^{LTR})$ & 94.3\% & 90.1\% \\
\hline
PIO$_0(K^{LTR})$ \& PIR$_1(K^{LTR})$ & 92.6\% & 90.1\% \\
\hline
PIR$_0(K^{RTL})$ \& PIR$_1(K^{LTR})$ & 92.6\% & 89.0\% \\
\hline
PIR$_0(K^{LTR})$ \& PIR$_0(K^{RTL})$ & 90.2\% & 89.0\% \\
\hline
PIO$_0(K^{LTR})$ \& PIR$_0(K^{RTL})$ & 91.9\% & 88.4\% \\
\hline
PIO$_0(K^{RTL})$ \& PIO$_1(K^{LTR})$ & 86.1\% & 86.0\% \\
\hline
PIO$_0(K^{BTT})$ \& PIO$_1(K^{LTR})$ & 84.2\% & 85.1\% \\
\hline
PIO$_1(K^{LTR})$ \& PIR$_1(K^{BTT})$ & 85.6\% & 84.8\% \\
\hline
PIO$_1(K^{LTR})$ \& PIR$_1(K^{TTB})$ & 83.1\% & 84.8\% \\
\hline
PIO$_0(K^{TTB})$ \& PIO$_1(K^{LTR})$ & 84.9\% & 84.8\% \\
\hline
PIO$_0(K^{BTT})$ \& PIO$_1(K^{RTL})$ & 85.6\% & 84.8\% \\
\hline 
    
    \end{tabular}
\caption{Summary of the 15 sweeping plane double descriptor vectors that achieved the highest OOS accuracy scors.} 
    \label{tab:tab1}
\end{table}

We applied $k-$means clustering to all sweeping plane double descriptor vectors and computed the average OOS accuracy for a range of values of $k$\new{, also known as an elbow plot} (S\ref{fig:SIfig4} Fig). 
While the average OOS accuracy decreased with increasing values of $k$, the rate of decrease slowed markedly for $k>5$. 
We fixed $k=5$ as this value robustly clustered the $(\rho,\chi)$ parameter space into biologically interpretable clusters (Fig~\ref{fig:fig13}).
Larger values of $k$ may overfit the data \cite{bholowalia_ebk-means_2014}, while smaller values 
fail to partition ($\rho$,$\chi$) parameter space into biologically interpretable regions (S\ref{fig:SIfig5} Fig). 
For example, when $k=3$, simulations with strong haptotaxis ($\rho>\chi$), equal chemotaxis and haptotaxis ($\rho=\chi$), and strong chemotaxis ($\rho<\chi$) are clustered together. When $k=4$, there are two distinct clusters near $\rho=\chi$, but simulations with large values of the haptotaxis parameter, $\rho$, are clustered with those with high values of the chemotaxis parameter, $\chi$. With $k=5$ clusters, the region with large values of the haptotaxis parameter, $\rho$, was further subdivided. We suggest that clustering with $k=5$ provides sufficient resolution to distinguish model simulations generated from different regions of the parameter space. \new{We note that more sophisticated descriptors could be used to justify this choice of $k$, including the silhouette score \cite{rousseeuw_silhouettes_1987}. 
%However, the focus of our study is to develop a pipeline that can be guided by expert knowledge, and we find that $k=5$ appears well justified by both our elbow plot as well as biological interpretation.
}

\begin{figure}
    \centering
    \includegraphics[width=0.99\textwidth]{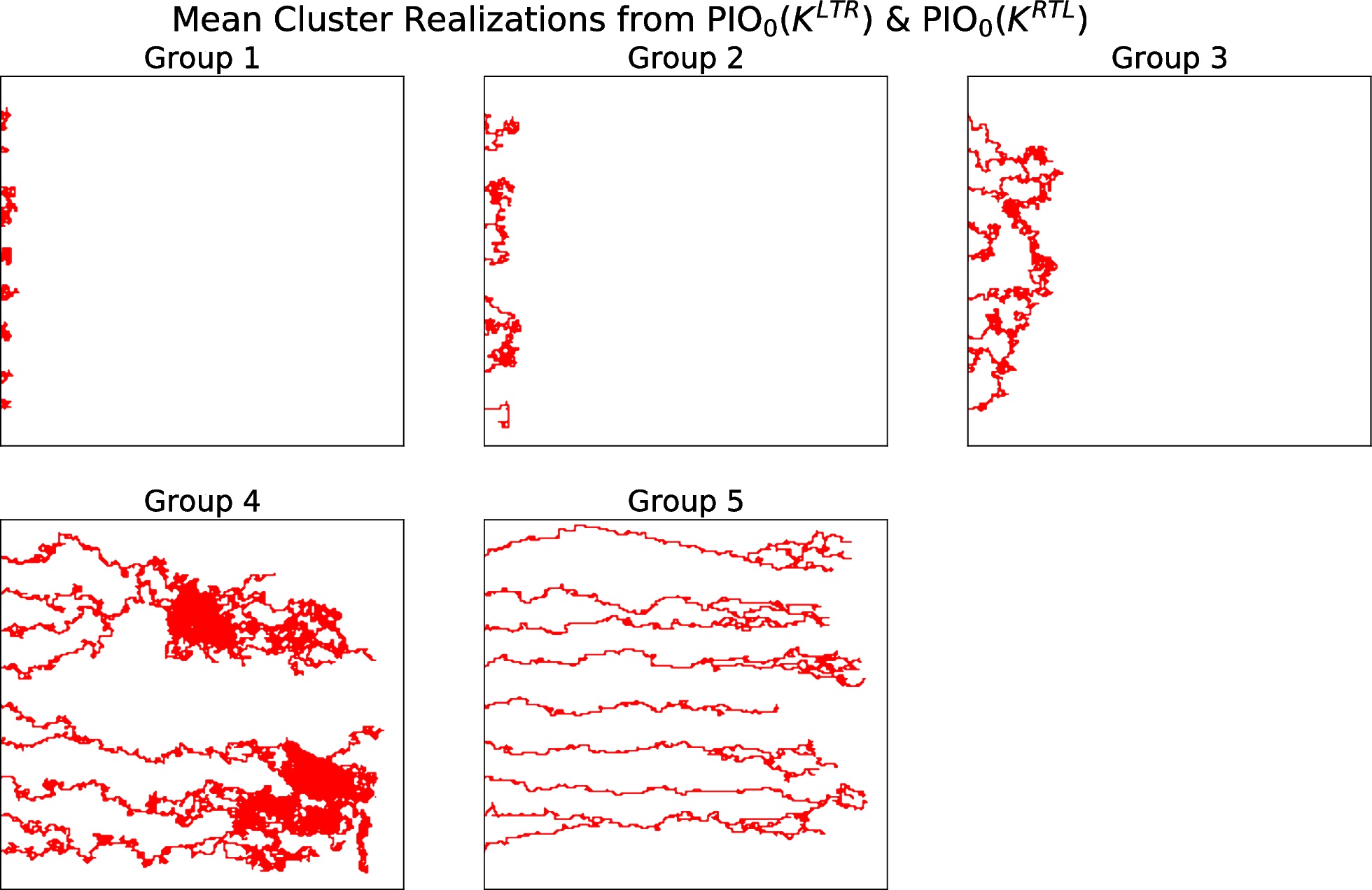}
    \caption{\textbf{Representative simulated blood vasculature from Groups 1-5.} Series of simulations that resemble the means from each of the five groups (as identified from the double descriptor vectors associated with the sweeping plane filtrations PIO$_0(K^{LTR})$ \& and PIO$_0(K^{RTL})$). The representative simulations for groups 1-5 were generated using the following pairs of parameter values: $(\rho,\chi) = (0.25,0), (0.5,0.2), (0.3,0.2), (0.15,0.15), \text{ and }(0.2,0.4)$, respectively.} 
    \label{fig:fig14}
\end{figure}

We identified five model simulations whose double descriptor vectors closely resembled the means from the clusters associated with the ``PIO$_0(K^{LTR})$ \& PIO$_0(K^{RTL})$" clustering.
%; recall that this clustering is biologically interpretable and robust. 
Fig~\ref{fig:fig14} presents representative simulations of each cluster. We note that as $\rho$ and $\chi$ vary, the spatial extent and connectedness of the network vary as expected (i.e., greater spatial extent in the $x$-direction as $\chi$ increases, and greater connectedness as $\rho$ increases). We applied a similar analysis to model simulations sampled from $(\rho,\chi, \psi)$ parameter space, where the non-negative parameter $\psi$ represents the rate at which tip cells branch. The resulting parameter clustering changed only with $\rho$ and $\chi$, suggesting that the vascular morphology does not depend on $\psi$ (S\ref{fig:SIfig6} Fig).

\section*{Conclusions and Discussion}

We have introduced a topological pipeline to analyze simulated data of tumor-induced angiogenesis \cite{anderson_continuous_1998}. We investigated different filtrations to feed into the persistent homology algorithm. We demonstrated that simple data science algorithms can group together topologically similar simulated data. Pairs of topological descriptor vectors generated from sweeping plane filtrations robustly clustered simulations into five biologically interpretable groups. Simulations within each group were
generated from similar values of the model parameters $\rho$ and $\chi$, which determine the sensitivity of the tip cells to haptotaxis and chemotaxis. 

The Betti numbers quantify connectedness of the vasculature across spatial scales. We computed PH of four sweeping plane filtrations (LTR, RTL, TTB, BTT), which quantify topological features as a plane moves across the simulated data in a particular direction. We found that the filtration direction provides information about different aspects of the vascular network. For the LTR filtration, the number of connected components (i.e., $\beta_0$) decreases when an anastomosis event occurs, whereas for the RTL filtration, $\beta_0$ decreases when a branching event occurs. The TTB and BTT filtrations summarize vascular tortuosity: when tip cell movement in the $y$-spatial direction is negligible and/or tip cells do not merge with adjacent sprouts, $\beta_0$ increases with the number of tip cells/ number of sprouts. 
We found that the combined PIO$_0(K^{LTR})$ \& PIO$_0(K^{RTL})$ sweeping plane descriptor vectors robustly cluster haptotaxis-chemotaxis parameter space into five distinct regions that appear to be connected. 

Our results may depend on the model setup, in which new blood vessels are created as endothelial tip cells move from left to right. In the future, we may explore the use of extended persistence \cite{Edelsbrunner2008} and/or PH transform \cite{turner2014persistent}, which allow consideration of multiple different directions in the topological analysis. 
If the vessel network grows radially, for example, then radial filtrations may be more informative than the LTR filtration \cite{Byrne2019,Stolz2020}. Classically in PH, topological features of short persistence are topological noise, yet there is growing evidence that these short features can provide information about geometry and curvature \cite{bendich_persistent_2016,Stolz2020,feng_persistent_2019,marchese_signal_2018,robins_principal_2016,stolz_persistent_2017,xia_persistent_2014,bubenik2020persistent}. 
Our results using persistence images with unit weightings, rather than ramped weightings, also suggest that short persistence features may be informative for distinguishing vessel morphologies. \new{The spatial resolution of the output binary images that we used for our analyses was the same as that used to generate the simulations, 
enabling us to distinguish the presence and absence of individual endothelial cells. In future work, we will consider finer and coarser  binary images to investigate how the performance of the methods depends on their spatial resolution.}

%\new{To cite when discussing efficiency of the TDA filtrations in 3d \cite{emblem_vessel_2014}.}
Many models of angiogenesis have been developed and implemented since the Anderson-Chaplain model, as has been extensively reviewed previously \cite{byrne_dissecting_2010,hadjicharalambous_tumour_2021,hamis_blackboard_2019,karolak_towards_2018,mantzaris_mathematical_2004,metzcar_review_2019,peirce_computational_2008,rockne_2019_2019,scianna_review_2013}. We plan to use the methods described in this work to interpret more sophisticated models of angiogenesis and, in the longer term, to determine how vessel morphology affects tumor survival and to predict the efficacy of vascular targetting agents. The tumor is not explicitly modeled in the Anderson-Chaplain model, but many studies couple angiogenesis and tumor growth \cite{owen_angiogenesis_2008,perfahl_3d_2017,norton_modeling_2018,vavourakis_validated_2017,vilanova_mathematical_2017}. 
For example, Vavourakis et al. developed a 3D model of angiogenic tumor growth that incorporates the effects of mechanotaxis on vessel sprouting and mechano-sensitive vascular remodelling \cite{vavourakis_validated_2017}. \new{Even though the Anderson-Chaplain model neglects mechanotaxis, subcellular signalling, and many other biophysical processes, it has been widely adopted due to its ability to generate networks which are in good agreement with those seen \emph{in vivo}.  More complex models have been developed that account for mechanical factors such as mechanotaxis, pressure-driven convective transport of extracellular factors, mechanical stimulation of endothelial cell proliferation, and subcellular signalling pathways that guide cell fate specification of tip and stalk cells \cite{vavourakis_validated_2017, vilanova_computational_2018, perfahl_3d_2017, stepanova_multiscale_2021}. Importantly, these more complex models have been shown to recapitulate experimental observations of time-varying spatial distributions of angiogenic vasculature \cite{vavourakis_validated_2017} and features such as asymmetric neovasculature \cite{vilanova_computational_2018} and cell rearrangements and phenotype switching of endothelial stalk and tip cells \cite{chen_endothelial_2019,jakobsson_endothelial_2010}. In future work we will apply the TDA framework developed here to these, and other angiogenesis models.
%: since our framework only relies on a segmented vascular image, this should, in principle, be relatively straightforward. 
Such analyses will enable us to investigate the effect that processes such as mechanotaxis and subcellular signalling have on the topology of blood vessel networks. }
\old{Vilanova et al. developed a phase-field model of vascular growth, regression, and re-growth that agrees with \emph{in vivo} experimental data \cite{vilanova_mathematical_2017}.}

Miller at al. used computational models to analyze tumor regression in response to drug treatment as a function of vasculature morphology and heterogeneity \cite{ha_evaluation_2018}. \old{Our future analysis of such state-of-the-art modeling techniques will require us to extend the topological methods presented here to handle heterogeneous data comprised of multiple cell types in 3D.} \new{Our future analysis of such state-of-the-art modeling techniques will require us to extend the topological methods presented here for spatio-temporal data that describes how multiple cell types (e.g., tumor cells, immune cells and endothelial cells) interact in 3D. In a preliminary study, we applied these methods to 3D data from \emph{in vivo} studies of tumor vessel networks, which is near the edge of current computational feasibility \cite{Stolz2020}. Further work in this direction}\old{. and would } requires significant computational advancements to scale the TDA pipeline to multiple, more complex models and analyze the parameter landscape.

 The topological approaches considered in this work represent a multiscale way to summarize the properties of \textit{in silico} vasculature, and could be extended to experimental image data. Previous studies have quantified experimental data of tumor-induced angiogenesis using standard descriptors that include sprout length and diameter, number of tumor cells, concentration of angiogenic chemicals
 \cite{feleszko_lovastatin_1999,cai_long_2017,sailem2020morphological}. To compare models and data will require descriptors, such as the topological ones proposed here, that can discriminate between different conditions (e.g., parameter values).
 The challenges of making direct comparisons between experimental and synthetic data is a significant bottleneck in the validation of mathematical and computational models of angiogenesis. In future work we will apply the topological approaches outlined in this paper to compare different model simulations with experimental data  \cite{gimbrone_tumor_1974,hlushchuk_morphological_2016}. In this way, we aim to determine biologically realistic parameter values for control conditions \cite{bauer_cell-based_2007}, and to identify parameter values that
 are associated with response to treatments, including vessel normalization.

\section*{Acknowledgments}
HAH, BJS and HB are grateful to the support provided by the UK Centre for Topological Data Analysis EPSRC grant EP/R018472/1. HAH gratefully acknowledges funding from EPSRC EP/R005125/1 and EP/T001968/1, the Royal Society RGF$\backslash$EA$\backslash$201074 and UF150238, and Emerson Collective. This work was partially supported by the National Science Foundation under Grant DMS-1638521 to the Statistical and Applied Mathematical Sciences Institute and IOS-1838314 to KBF, and in part by National Institute of Aging grant R21AG059099 to KBF. The funders had no role in study design, data collection and analysis, decision to publish, or preparation of the manuscript.

\appendix

\subsection*{Modeling of Tumor-induced angiogenesis}

\subsubsection*{Model Equations for biological movement and reaction}

We use the Anderson-Chaplain model \cite{anderson_continuous_1998} to simulate the formation of blood vessel networks in response to tumor-generated chemical gradients, as depicted in Fig 1. This model describes the spatio-temporal evolution of three dependent variables, namely endothelial tip cells, tumor angiogenic factor (TAF), and fibronectin. Endothelial tip cells emerge from a parent blood vessel and migrate towards the tumor in response to spatial gradients of TAF. In practice, tumors secrete multiple TAFs, including vascular endothelial growth factor (VEGF) and basic fibroblast growth factor (bFGF); in the Anderson-Chaplain model, TAF can be viewed as a generic factor. Tip cells migrate by chemotaxis up spatial gradients and also exhibit a small amount of random diffusion. Fibronectin, a major component of the extracellular matrix, also influences tip cell movement; 
tip cells move, via haptotaxis, up spatial gradients of fibronectin. For a 2D Cartesian geometry, the evolution of the tip cell density can be represented by the following continuous partial differential equation:
\begin{equation}
\dfrac{\partial n}{\partial t}	=D_n\Delta n-\chi_0 \nabla\cdot\left(n\nabla c\right)-\nabla\cdot(\rho_0 n\nabla f) \label{eq:n_dynamics}
\end{equation}
where $n=n(t,x,y)$ denotes tip cell density at time $t$ and spatial position $(x,y)$, $f=f(t,x,y)$ denotes fibronectin concentration, and $c=c(t,x,y)$ denotes TAF concentration. The parameters $D$, $\chi$, and $\rho$ parameterize tip cell diffusion, chemotaxis, and haptotaxis, respectively.

Fibronectin is assumed to be produced and consumed by tip cells and to remain bound to substrate ECM after its production. 
The evolution of fibronectin concentration can be represented by the following partial differential equation:
\begin{equation}
\dfrac{\partial f}{\partial t}	=\omega n-\mu nf, \label{eq:f_dynamics}    
\end{equation}
where the parameters $\omega$ and $\mu$ parameterize fibronectin production and consumption, respectively. 

TAF is assumed to be secreted by the tumor and to diffuse on a scale faster than the endothelial cell migration. We assume tip cells consume TAF while migrating towards the tumor. The evolution of TAF concentration can be represented by the following partial differential equation:
\begin{equation}
    \dfrac{\partial c}{\partial t}	= D_c\Delta c -\lambda nc, \label{eq:c_dynamics_int}
\end{equation}
where the parameters $D_c$ and $\lambda$ parameterize TAF diffusion and consumption by tip cells, respectively. Following \cite{anderson_continuous_1998}, we will replace Equation \eqref{eq:c_dynamics_int} 
with the simpler partial differential equation:
\begin{equation}
    \dfrac{\partial c}{\partial t}	= -\lambda nc. \label{eq:c_dynamics}
\end{equation}

The parameters that appear in the model equations \eqref{eq:n_dynamics}-\eqref{eq:f_dynamics} and \eqref{eq:c_dynamics} are listed in S1 Table, together with a brief description of their biological meaning and typical dimensional values taken from \cite{anderson_continuous_1998}.

\subsubsection*{Nondimensionalization, initial \& boundary conditions:} %%% Nondimensionalization goes here

Following \cite{anderson_continuous_1998}, we nondimensionalize Eqs. \eqref{eq:n_dynamics}--\eqref{eq:f_dynamics} and \eqref{eq:c_dynamics} as follows. We consider the nondimensionalized variables (with asterisks) given by 
\begin{align}
    t^* &= \dfrac{D_c}{L^2}t, \hspace{1cm} n^* = \dfrac{n}{n_0}, \hspace{1cm} c^*=\dfrac{c}{c_0}, \hspace{1cm} \rho^*=\dfrac{\rho}{\rho_0}, \nonumber \\
    D^* &= \dfrac{D_n}{D_c}, \hspace{1cm} \chi^* = \dfrac{\chi_0c_0}{D_c}, \hspace{1cm} \rho^* = \dfrac{\rho_0f_0}{D_c}, \nonumber \\
    \beta^* &= \dfrac{\omega L^2 n_0}{f_0 D_c}, \hspace{1cm} \gamma^* = \dfrac{\mu L^2 n_0}{D_c}, \hspace{1cm} \eta^* = \dfrac{\lambda L^2 n_0}{D_c}. \label{eq:nondimensionalize}
\end{align}
We substitute the variables from Equation \eqref{eq:nondimensionalize} into Equations \eqref{eq:n_dynamics}--\eqref{eq:f_dynamics} and \eqref{eq:c_dynamics} and find the nondimensionalized system of equations given by (asterisks dropped for notational convenience)
\begin{align}
    \frac{\partial n}{\partial t } &= D\Delta n - \chi\nabla \cdot (n \nabla c ) - \rho \nabla (n \nabla f ),  \nonumber \\
    \frac{\partial f}{\partial t } &= \beta n - \gamma nf,  \nonumber \\
    \frac{\partial c}{\partial t } &= - \eta nc.  \label{eq:nondimensionalize_app}
\end{align}
We list the nondimensionalized parameters for Equation in S2 Table.

\begin{align}
    \frac{\partial n}{\partial t } &= D\Delta n - \chi\nabla \cdot (n \nabla c ) - \rho \nabla \cdot (n \nabla f )  \nonumber \\
    \frac{\partial f}{\partial t } &= \beta n - \gamma nf  \nonumber \\
    \frac{\partial c}{\partial t } &= - \eta nc   \label{eq:System_of_eqns}
\end{align}
The dimensionless parameters used in all simulations (unless otherwise noted) are listed in S2 Table.

We assume that the initial distribution of the TAF $c$ can be described as a one-dimensional profile from a tumor source at $x=1$ as
\begin{equation}
    c(x,y,0) = e^{-\frac{1-x^2}{\epsilon_1}}, (x,y)\in[0,1]\times[0,1].\label{eq:TAF_linear}  
\end{equation}
The initial distribution of the fibronectin concentration $f$  is given by
\begin{equation}
    f(x,y,0) = ke^{\frac{-x^2}{\epsilon_2}}, (x,y)\in[0,1]\times[0,1]. \label{eq:fibronectin}
\end{equation}
Equations \eqref{eq:TAF_linear} and \eqref{eq:fibronectin} are visually depicted in S7 Fig.

We use no-flux conditions at all boundary points to ensure that the net flux of tip cells into the spatial domain is zero (no boundary conditions are needed for fibronectin or TAF since their governing equations do not contain spatial derivatives).  

\subsubsection*{Stochastic ABM implementation:} We simulate Equation \eqref{eq:System_of_eqns} by using a stochastic agent-based model (ABM) to describe tip cell locations over time and the resulting concentrations of fibronectin and TAF. In turn, we create network of angiogenic blood vessels comprised of the tip cells and their previous locations. We begin simulations by creating a lattice for our spatial domain given by $\bm{x} = i\Delta x, i = 0,...,N, \Delta x = 1/N$ with $N=201$. The discretization of the second spatial dimension, $y$, is defined equivalently. We then initialize ten tip cells locations over time, $n_i^{\text{tip}}(t)$, by setting $n_i^\text{tip}(0) = \{(0,0.1i)\}_{i=1}^{9}$.

In order to simulate Eqs. \eqref{eq:n_dynamics}-\eqref{eq:f_dynamics} and \eqref{eq:c_dynamics}, we discretize these equations around each $n_i^\text{tip}(t)$ over timestep $\Delta t$ using the upwind method and form five probabilities governing whether the given tip cell will move left, right, up, down, or stay in place. We then set $n_i^\text{tip}(t+\Delta t)$ equal to one of these locations based on these five computed probabilities.  

We define the $i^\text{th}$ \new{vessel segment} as 
\begin{equation}
n_i^\text{\new{vessel}} = \bigcup_t n_i^\text{tip}(t).  
\end{equation}

We also consider tip cell branching, in which a new tip cell is placed in the ABM domain \new{and creates a new vessel segment}. \new{As in} \old{Similar to} \cite{anderson_continuous_1998}, we use the following three rules to determine when a tip cell branches and forms a new \new{vessel segment}:
\begin{enumerate}
    \item Only sprouts of age greater than some new threshold value, $\psi$, may branch. We set $\psi=0.1$. After a tip cell branches, its ``age'' is re-set to zero.
    \item  \new{A tip cell can only branch into neighboring lattice sites that are not occupied by another sprout.}
    \old{Tip cells can only branch if it neighbors lattice sites that are not occupied by any other sprouts.}
    \item The endothelial cell density at $n_i^\text{tip}(t)$ must be greater than some threshold level given by
    \begin{equation}
    n_b = \dfrac{2.5}{c(n_i^\text{tip}(t),t)}.
    \end{equation}
    We define \new{the} endothelial cell density at the point $(x,y)$ as the number of \new{locations occupied by endothelial cells} \old{occupied sprout locations} in the $3\times 3$ grid of points centered at $(x,y)$, divided by nine.
\end{enumerate} 
If all three of the above conditions are satisfied, then the tip cell will branch and randomly place the new tip cell  into one of its eight adjacent and unoccupied lattice sites.

\vspace{0.5cm}\hspace{-0.5cm}\textbf{Resulting vasculature image:} From an ABM simulation, we can form the entire vasculature by constructing
\begin{equation}
    \bm{N}_\text{vasculature} = \bigcup_i n_i^\text{sprout}. \label{eq:network}
\end{equation}
We further convert this vasculature to a binary image, $\bm{N}$, where we set each pixel $(i,j)$ to be
\begin{equation}
    \bm{N}(i,j) = \begin{cases} 1 & \text{if $(i\Delta x,j\Delta y) \in N_\text{network}$} \\ 0 & \text{otherwise} \\  \end{cases}. \label{eq:binary_images_Defn} 
\end{equation}
We present three representative realizations of $\bm{N}$ in Fig 7 that result from different $(\rho,\chi)$ combinations. In these figures, red cells have a values of one to denote a sprout, and white cells have a value of zero to denote the absence of a sprout.

\subsection*{Clustering results}

\textbf{Clustering with standard descriptor vectors:} The use of the standard biological descriptor vectors does not lead to biologically interpretable clusterings of the ($\rho$,$\chi$) parameter space (S1 Fig). The clustering of the parameter space using tips and vessel segment descriptor vectors suggested that the majority of simulations belong to one large cluster, which demonstrates that these descriptor vectors cannot distinguish between simulations that result from visually distinct vascular architectures generated by different model parameter values. The final vasculature length descriptor vector led to four distinct clusters in the parameter space, but almost all ($\rho$,$\chi$) pairs with $\chi>\rho$ are placed into one large cluster (group 5), suggesting that this descriptor vector cannot distinguish over half of the model simulations from each other. Concatenating all three descriptor vectors into one large descriptor vector led again to one dominant cluster. These descriptor vectors clusterings are robust, however, as each achieves a high OOS accuracy of above 90\% (S1 Fig).

\textbf{Clustering with topological flooding descriptor vectors:} We clustered our simulated model data set using the the topological flooding descriptor vectors. Each descriptor vector counts either the number of connected components ($\beta_0$) or loops ($\beta_1$); performs the flooding filtration; and is summarized using a BC, PIR, or PIO, so we considered $2\times1\times3=6$ separate flooding topological descriptor vectors. Each of these six descriptor vectors are ranked according to their OOS accuracy scores in S3 Table. These clustering regions did not appear biologically interpretable, as many of the groups are not well connected in the $(\rho,\chi)$ parameter space (S4 Fig). These flooding descriptor vector clusterings are not robust, as we found a highest OOS accuracy of 65.8\% is achieved with the PIR$_1(K^{flood})$ descriptor vector.

We next considered concatenating doubles of topological flooding vectors as a means to improve robustness. We considered all (6 choose 2) = 15 possible flooding descriptor vectors doubles and ranked these clusterings by the highest OOS accuracies in S4 Table. Doubles of flooding descriptor vectors attain only slightly higher OOS accuracy scores than individual flooding descriptor vectors. The PIO$_0(K^{flood})$ \&  PIR$_1(K^{flood})$ double achieved the highest OOS classification accuracy of 66.4\%. Though the clustering regions in $(\rho,\chi)$ space for this double flooding descriptor vector did appear biologically interpretable, as the five groups are mostly well connected, aside from Group 4 (S5 Fig).

\textbf{Clustering with individual topological sweeping plane descriptor vectors:} We clustered our simulated model dataset using the the topological sweeping plane descriptor vectors. Each descriptor vector counts either the number of connected components ($\beta_0$) or loops ($\beta_1$); sweeps the plane in one of four directions (LTR, RTL, TTB, and BTT); and is summarized using a Betti curve (BC), persistence image with ramp weighting (PIR), or persistence image with one weighting (PIO), so we computed $2\times4\times3=24$ sweeping plane descriptor vectors for each model simulation.  Each of these 24 summaries were ranked according to their OOS accuracy scores in S5 Table. The four highest OOS accuracy scores resulted from the PIR$_1(K^{LTR})$ (91.5\% OOS accuracy), PIO$_0(K^{LTR})$ (83.5\% OOS accuracy), PIO$_1(K^{RTL})$ (74.9\% OOS accuracy), and PIO$_1(K^{LTR})$ (74.7\% OOS accuracy) descriptor vectors. While these clusterings result in high classification scores, the resulting clusters were determined to not be biologically interpretable. For example, S2 Fig shows that the parameter clusterings that result from either the PIR$_1(K^{LTR})$ or the PIO$_0(K^{LTR})$ descriptor vector include both high haptotaxis (\emph{e.g.}, ($\rho,\chi$)=(0.5,0)) and high chemotaxis (\emph{e.g.}, ($\rho,\chi$)=(0,0.5)) parameter combinations in the same cluster (group 5), even though these parameter combinations lead to markedly different vascular morphologies. The PIO$_1(K^{RTL})$ and PIO$_1(K^{LTR})$ clusterings also led to clusterings that include high chemotaxis and high haptotaxis simulations within the same cluster (within groups 1 and 2, respectively).

%\nolinenumbers

% Either type in your references using
% \begin{thebibliography}{}
% Text
% \end{thebibliography}
%
% or
%
% Compile your BiBTeX database using our plos2015.bst
% style file and paste the contents of your .bbl file
% here. See http://journals.plos.org/plosone/s/latex for 

%\bibliographystyle{abbrv}      % mathematics and 
%\bibliography{references.bib}  

\bibliographystyle{abbrv}      % mathematics and 
\bibliography{references.bib}  

\section*{Supporting information}

\begin{figure}
    \centering
    \includegraphics[width=.75\textwidth]{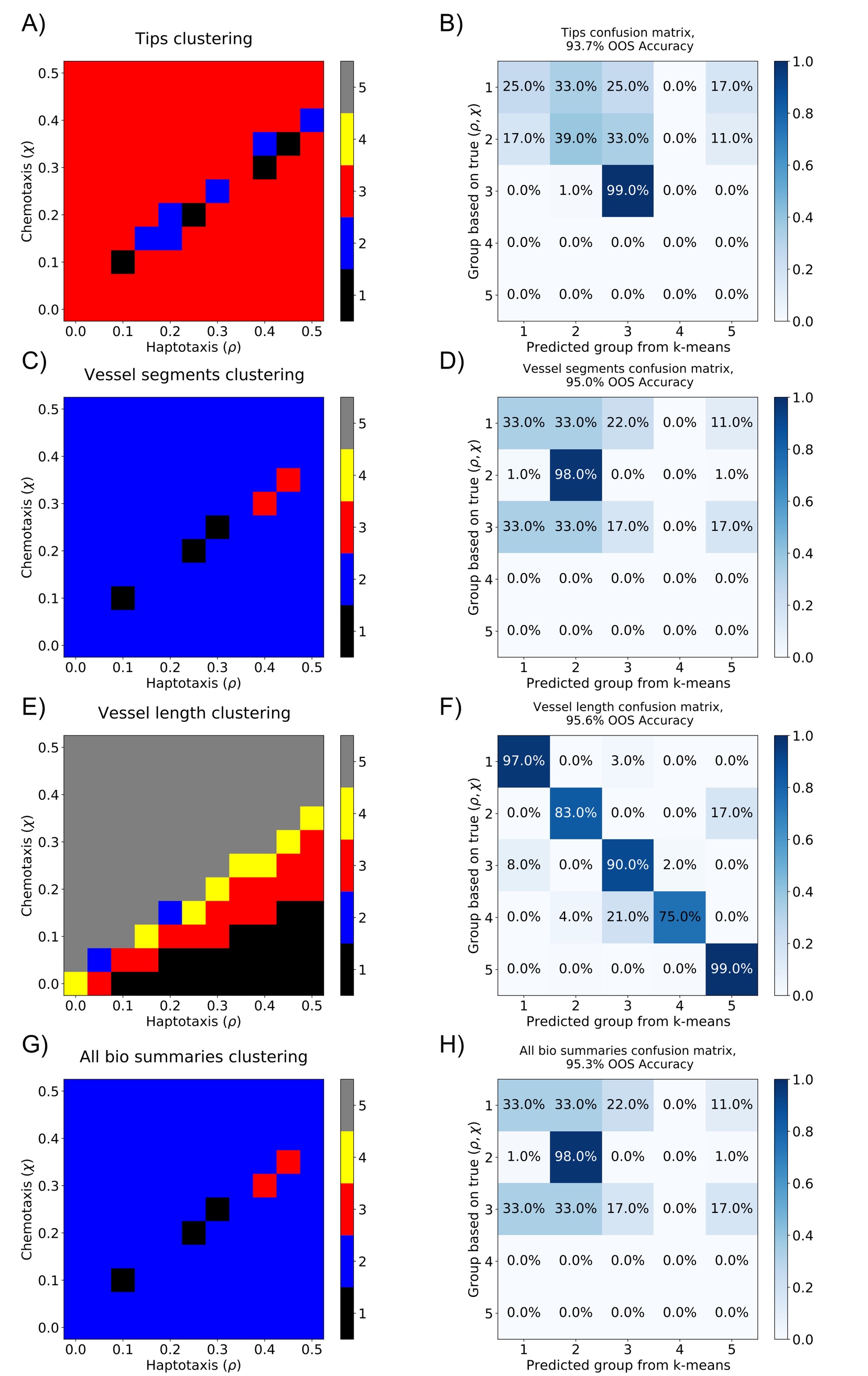}
    \caption{\textbf{Standard descriptor vector clustering.} Clustering of the $(\rho,\chi)$ parameter space using $k$-means with $k=5$ on the standard descriptor vectors to summarize each simulated vasculature. We considered A) Number of tips over time, C) Number of vessel segments over time, E) Final length of the simulated vasculature, and G) All of the previous descriptor vectors concatenated into one descriptor vector. The five clusters are ordered according to the mean $\chi$ value within the cluster. Panels B,D,F,H depict the out of sample confusion matrices for each descriptor vector.}
    \label{fig:SIfig1}
\end{figure}

\begin{figure}
    \centering
    \includegraphics[width=.75\textwidth]{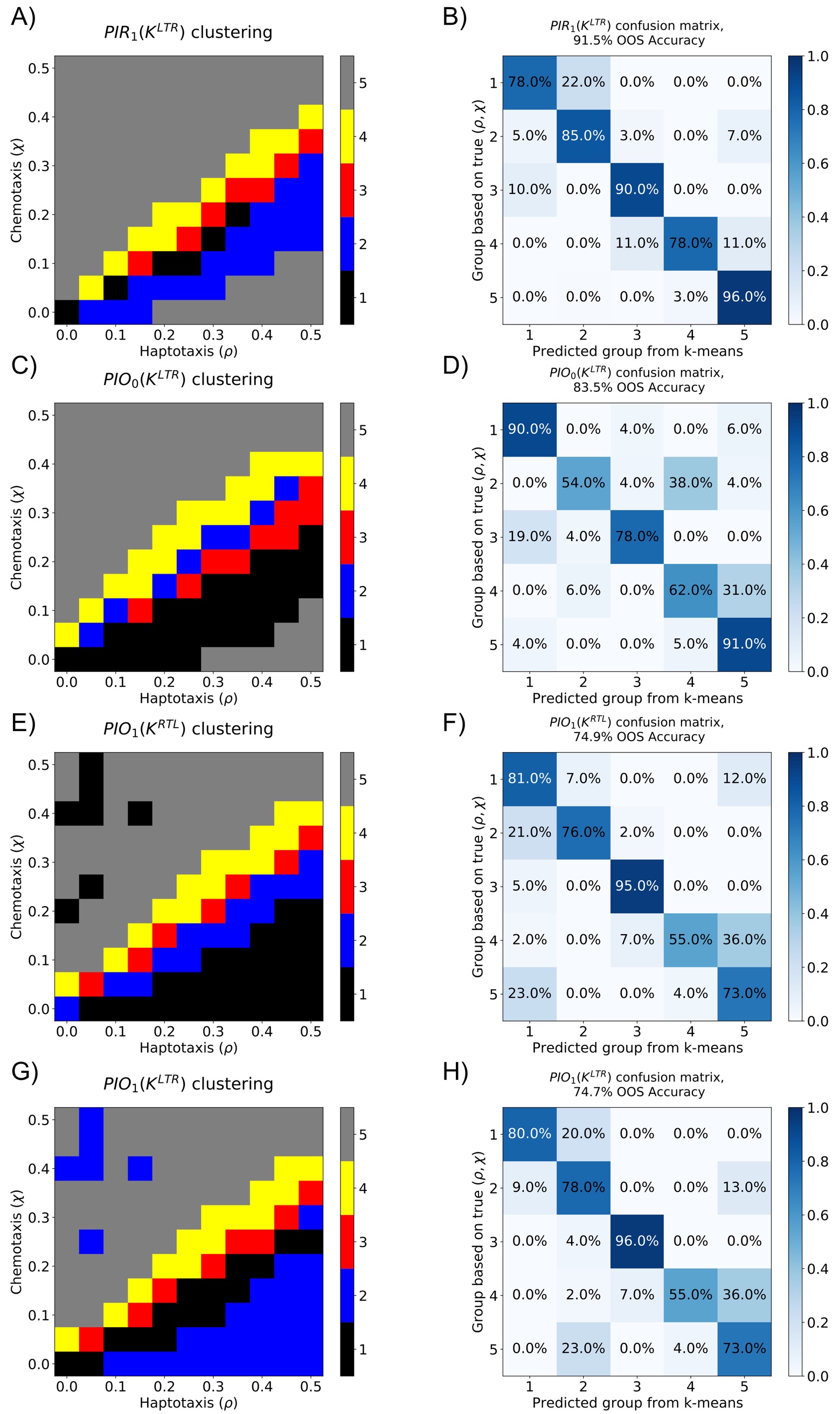}
    
    \caption{\textbf{Individual plane sweeping descriptor vector clustering.} Clustering of the $(\rho,\chi)$ parameter space using $k$-means with $k=5$ on individual sweeping plane topological filtrations to summarize each simulated vasculature. The four highest OOS accuracies resulted from the A) $PIR_1(K^{LTR})$, C) $PIO_0(K^{LTR})$, E) $PIO_1(K^{RTL})$, G) and $PIO_1(K^{LTR})$ descriptor vectors. The five clusters are ordered according to the mean $\chi$ value within the cluster. Panels B,D,F,H depict the out of sample confusion matrices for each descriptor vector.}
    \label{fig:SIfig2}
\end{figure}

\begin{figure}
    \centering
    \includegraphics[width=.5\textwidth]{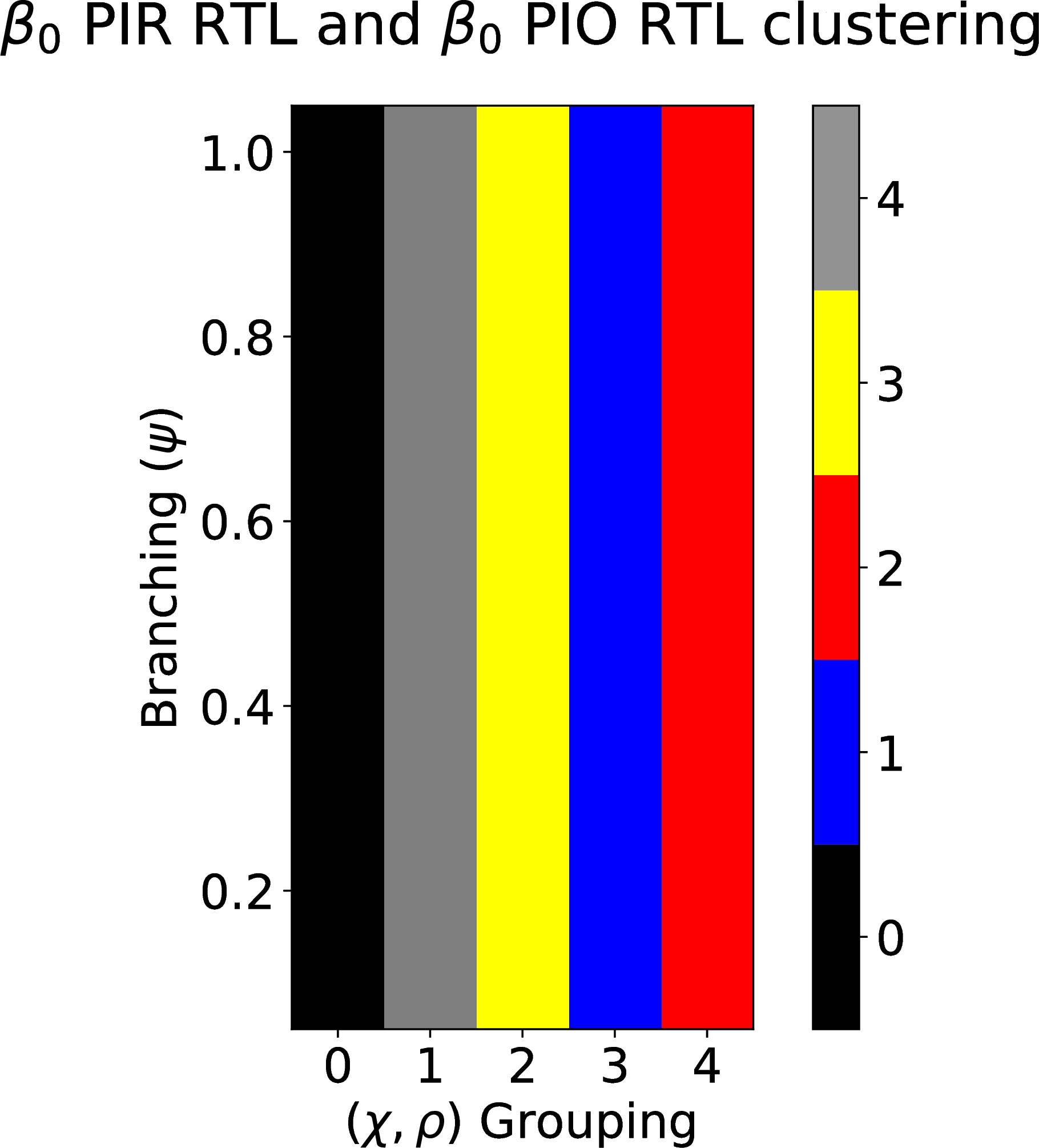}
    \caption{\textbf{Dimensionality reduction.} Dimensionality reduction of the PIO$_0(K^{LTR})$ \& PIO$_0(K^{RTL})$ double descriptor vector. We reduced the dimensionality of the PIO$_0(K^{LTR})$ \& PIO$_0(K^{RTL})$ double descriptor vector to two dimensions using principal components analysis and plot the reduced-dimension descriptor vector for each simulation. The color of each dot denotes the predicted grouping of each simulation from the $k$-means algorithm.}
    \label{fig:SIfig3}
\end{figure}

\begin{figure}
    \centering
    \includegraphics[width=.5\textwidth]{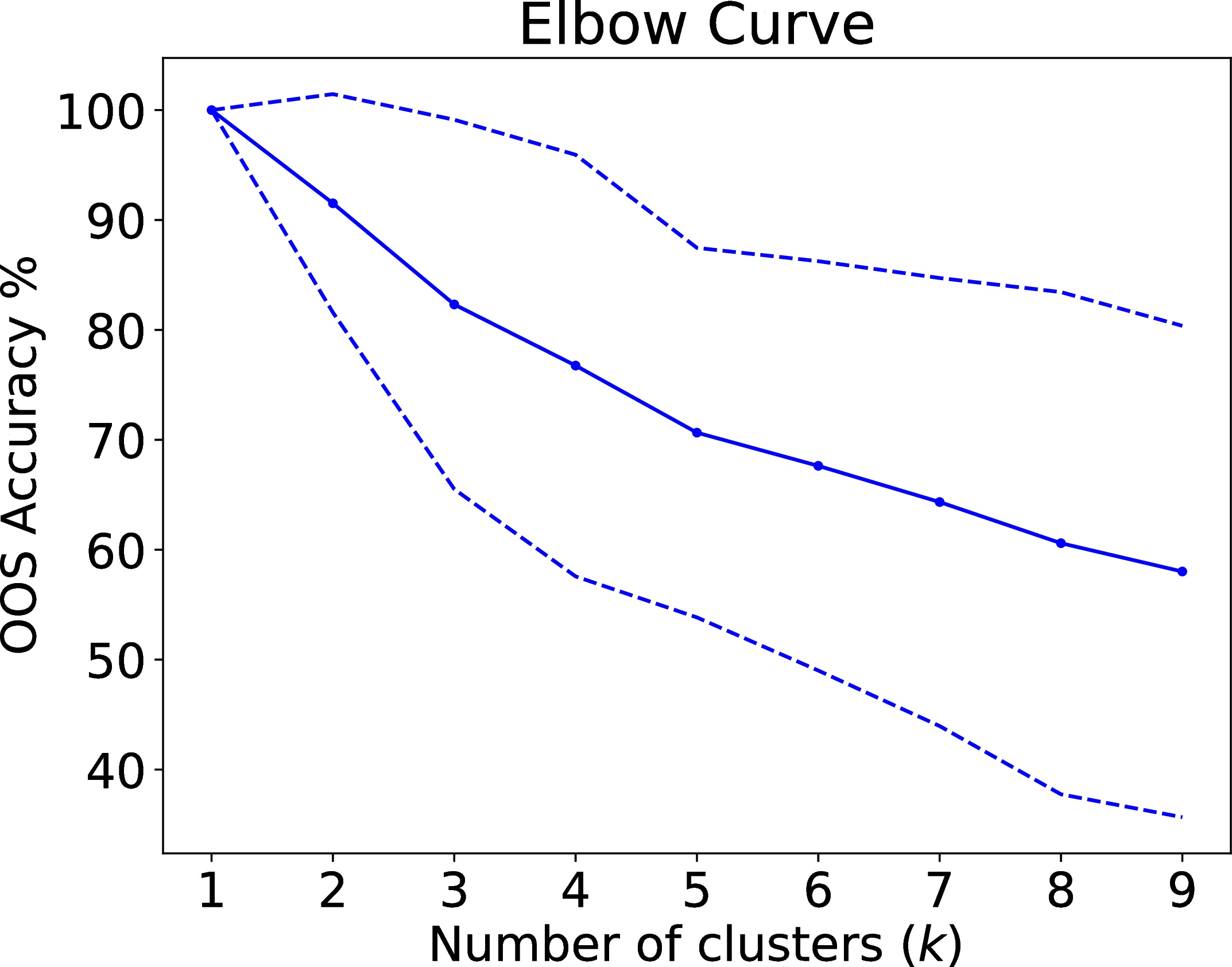}
    \caption{\textbf{Elbow Curve.} Elbow curve for the OOS Accuracy for the $k$-means clustering and labeling algorithm over different values of $k$. We considered all 276 doubles of topological feature vectors and plot the mean OOS accuracy percentage plus or minus two standard deviations for multiple values of $k$. We propose that the elbow occurs at $k=5$.}
    \label{fig:SIfig4}
\end{figure}

\begin{figure}
    \centering
    \includegraphics[width=\textwidth]{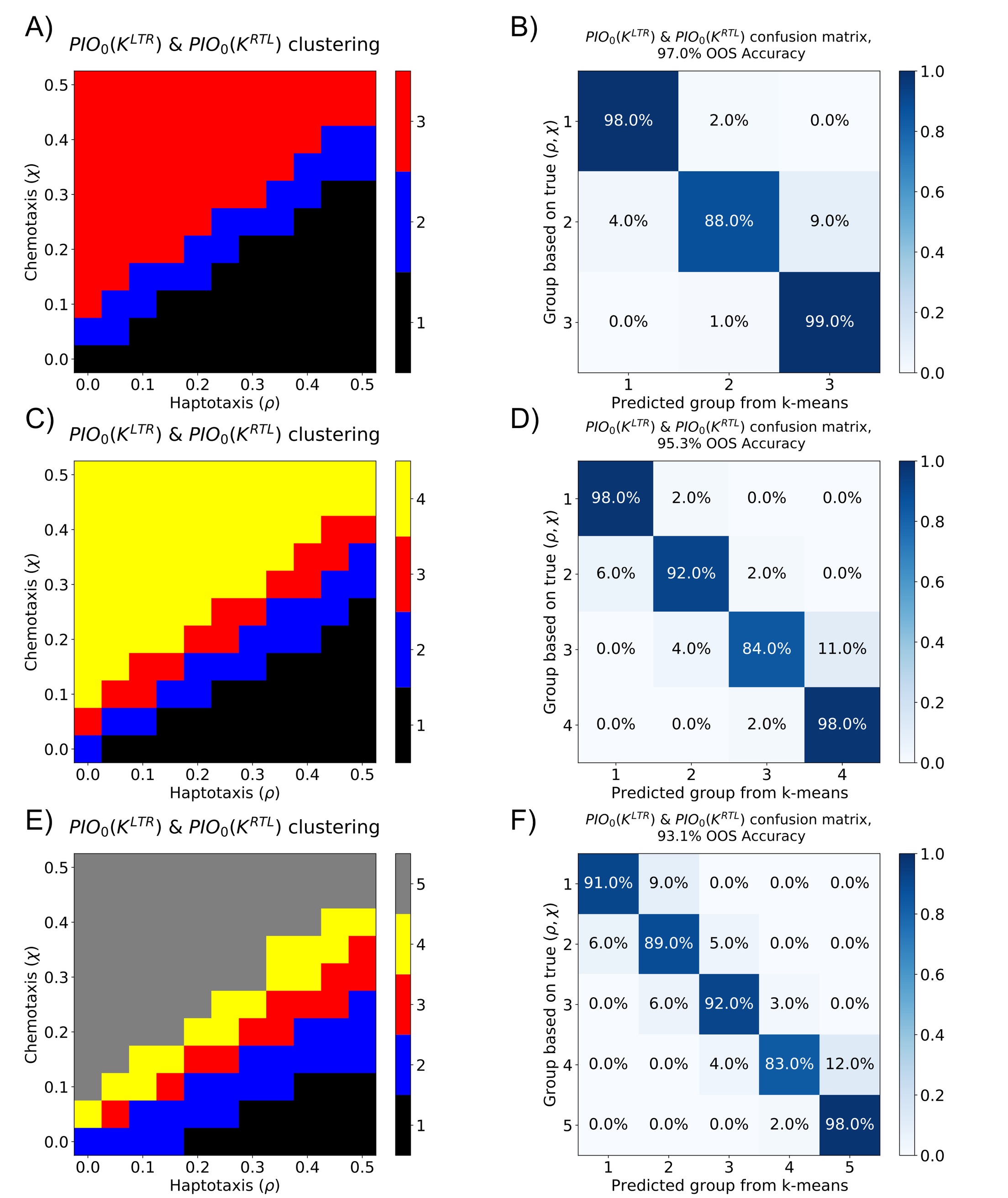}
    
    \caption{\textbf{Multiple $k$-means clusterings.} Clustering of the $(\rho,\chi)$ parameter space using the PIO$_0(K^{LTR})$ \& PIO$_0(K^{RTL})$ double of topological descriptor vectors using the $k$-means algorithm with A) $k=3$, C) $k=4$, E) $k=5$. The clusters are ordered according to the mean $\chi$ value within the cluster. Panels B,D,F) depict the out of sample confusion matrices for each descriptor vectory.}
    \label{fig:SIfig5}
\end{figure}

\begin{figure}
    \centering
    \includegraphics[width=.5\textwidth]{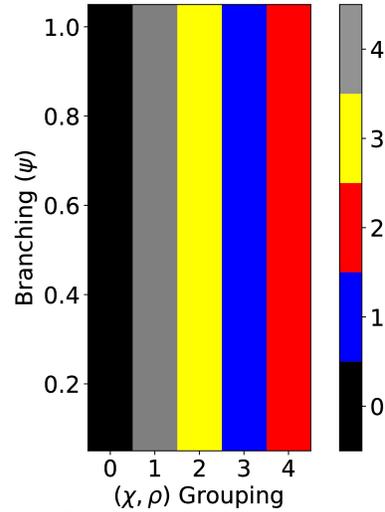}
    \caption{\textbf{$(\rho,\chi,\psi)$ clustering.}  Clustering based on the $(\rho,\chi)$ grouping and the time to branch, $\psi$. We created a data set of blood vessel simulations by considering ($\rho,\chi$) values from groups 1-5 from Fig~\ref{fig:fig13}A and letting $\psi$ vary over the 10 values $\{.1,0.2,0.3,\dots,1.0\}$. We simulated the Anderson-Chaplain model ten times at each ($\rho,\chi,\psi$) values to create 500 total simulations. We performed our clustering methodology on this collection of images based off the PIO$_0(K^{LTR})$ \& PIO$_0(K^{RTL})$ descriptor vector.  }
    \label{fig:SIfig6}
\end{figure}

\begin{figure}
    \centering
    \includegraphics[width=\textwidth]{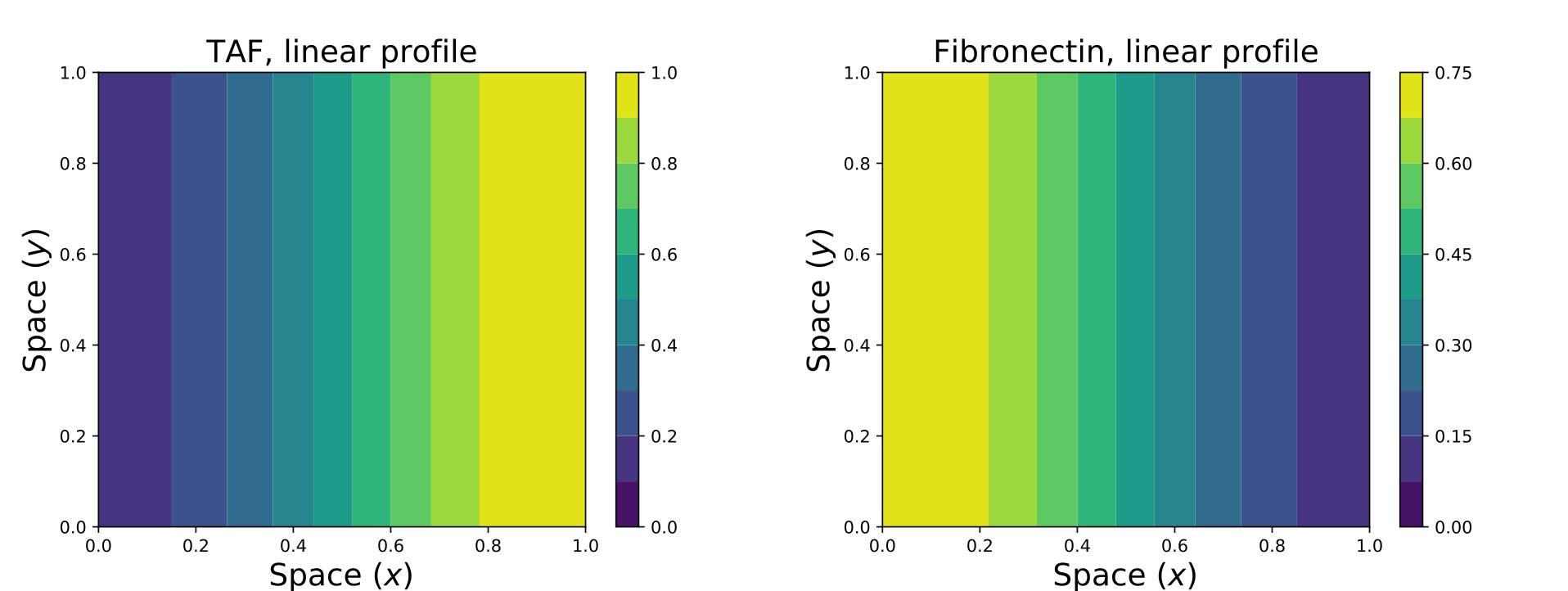}
    \caption{\textbf{Initial TAF and Fibronectin profiles.} TAF concentration (left) and fibronectin concentration (right).}
    \label{fig:SIfig7}
\end{figure}

\begin{figure}
    \centering
    \includegraphics[width=.75\textwidth]{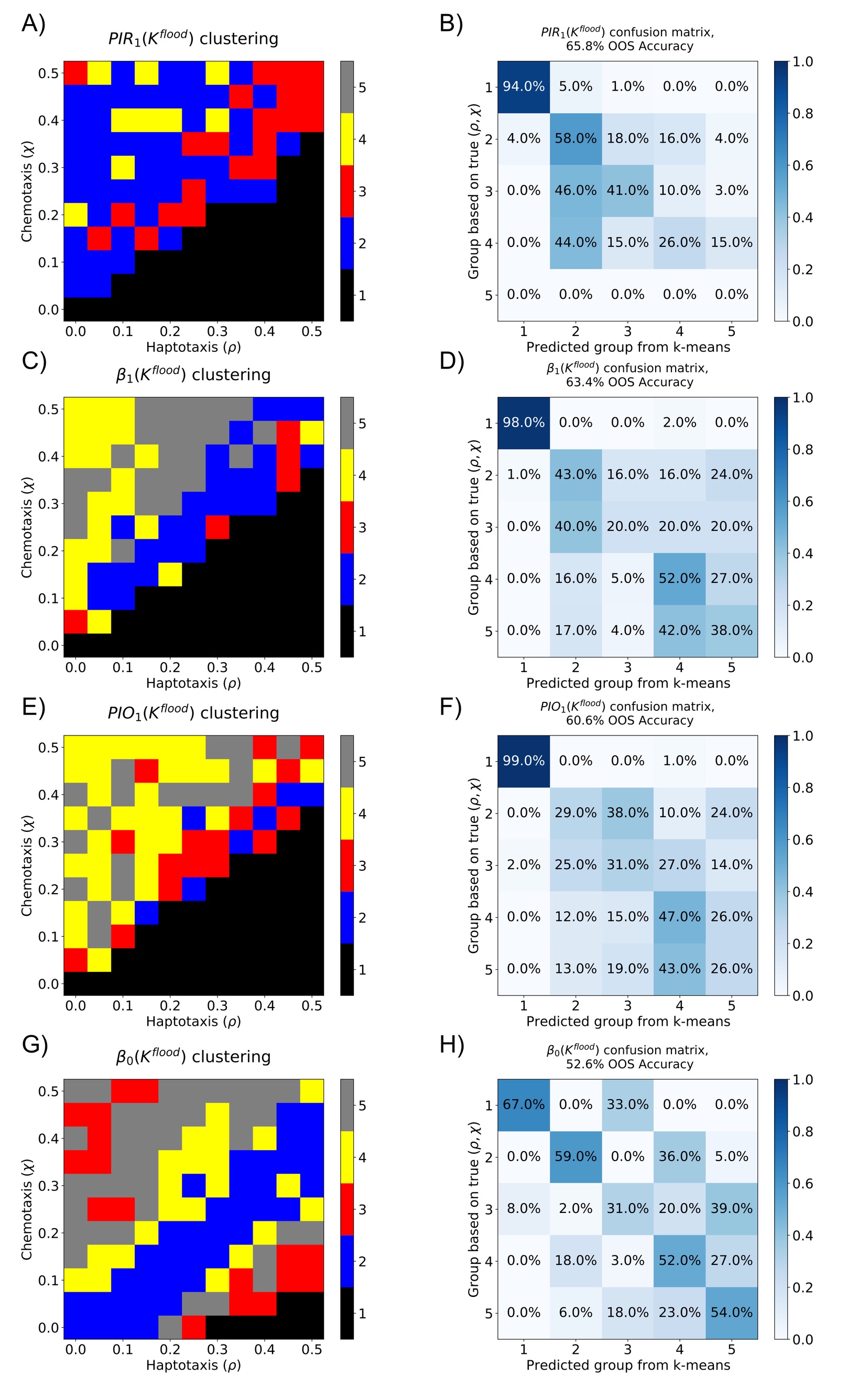}
    
    \caption{\textbf{Individual flooding descriptor vector clustering.} Clustering of the $(\rho,\chi)$ parameter space using $k$-means with $k=5$ on individual flooding topological descriptor vectors to summarize each simulated vasculature. The four highest OOS accuracies resulted from the A) $PIR_1(K^{flood})$, C) $\beta_1(K^{flood})$, E) $PIO_1(K^{flood})$, G) and $\beta_0(K^{flood})$ descriptor vectors. The five clusters are ordered according to the mean $\chi$ value within the cluster.  Panels B,D,F,H depict the out of sample confusion matrices for each descriptor vector.}
    \label{fig:SIfig8}
\end{figure}

\begin{figure}
    \centering
    \includegraphics[width=.75\textwidth]{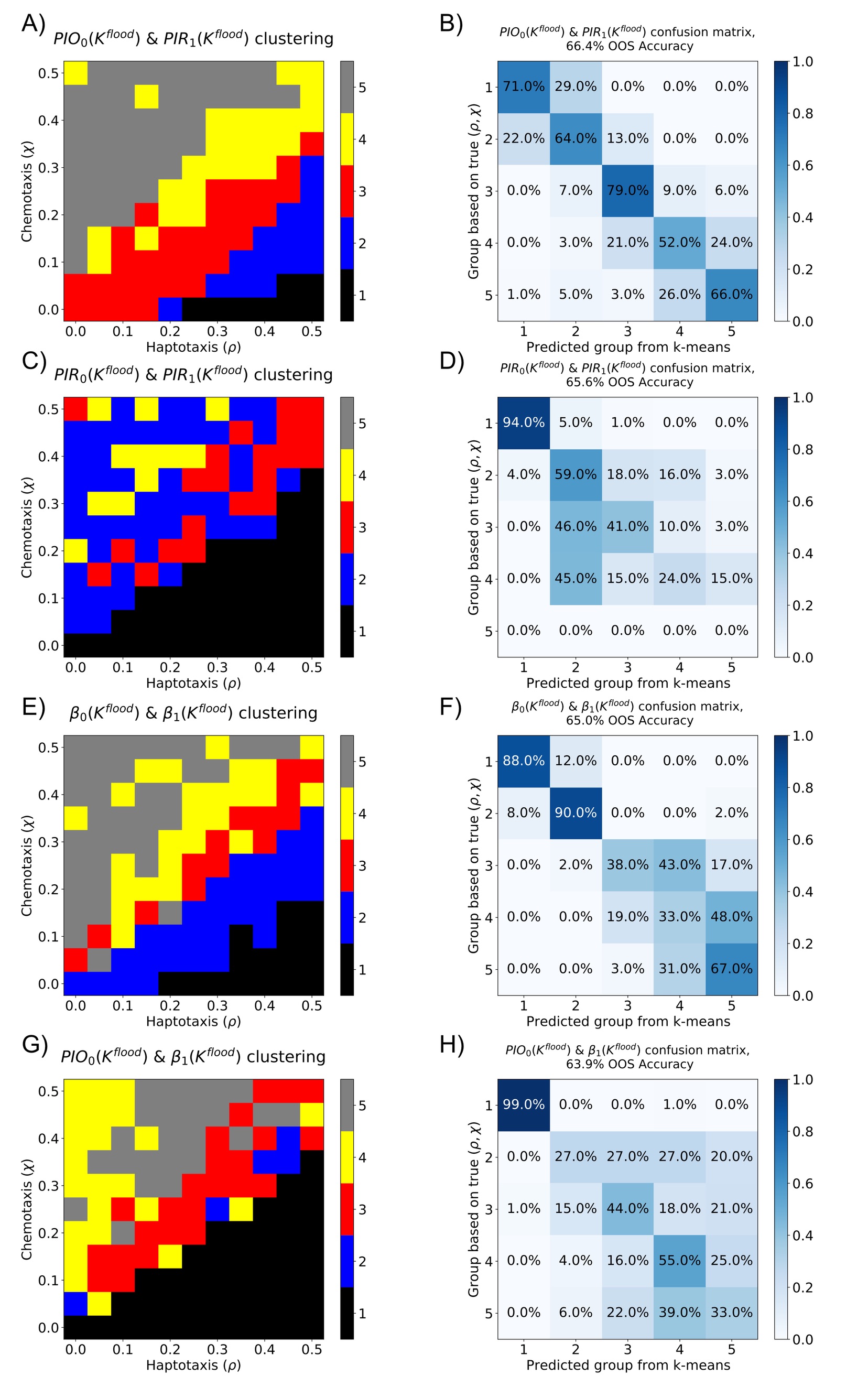}
    \caption{\textbf{Double flooding descriptor vector clustering.} Clustering of the $(\rho,\chi)$ parameter space using $k$-means with $k=5$ on double flooding topological descriptor vectors to summarize each simulation. The four highest OOS accuracies resulted from the A) $PIO_0(K^{flood}) \& PIR_1(K^{flood})$, C) $PIR_0(K^{flood}) \& PIR_1(K^{flood})$, E) $\beta_0(K^{flood}) \& \beta_1(K^{flood})$, G) and $PIO_0(K^{flood}) \& \beta_1(K^{flood})$ descriptor vectors. The five clusters are ordered according to the mean $\chi$ value within the cluster.  Panels B,D,F,H depict the out of sample confusion matrices for each descriptor vector.}
    \label{fig:SIfig9}
\end{figure}

\begin{figure}
    \centering
    \includegraphics[width=.75\textwidth]{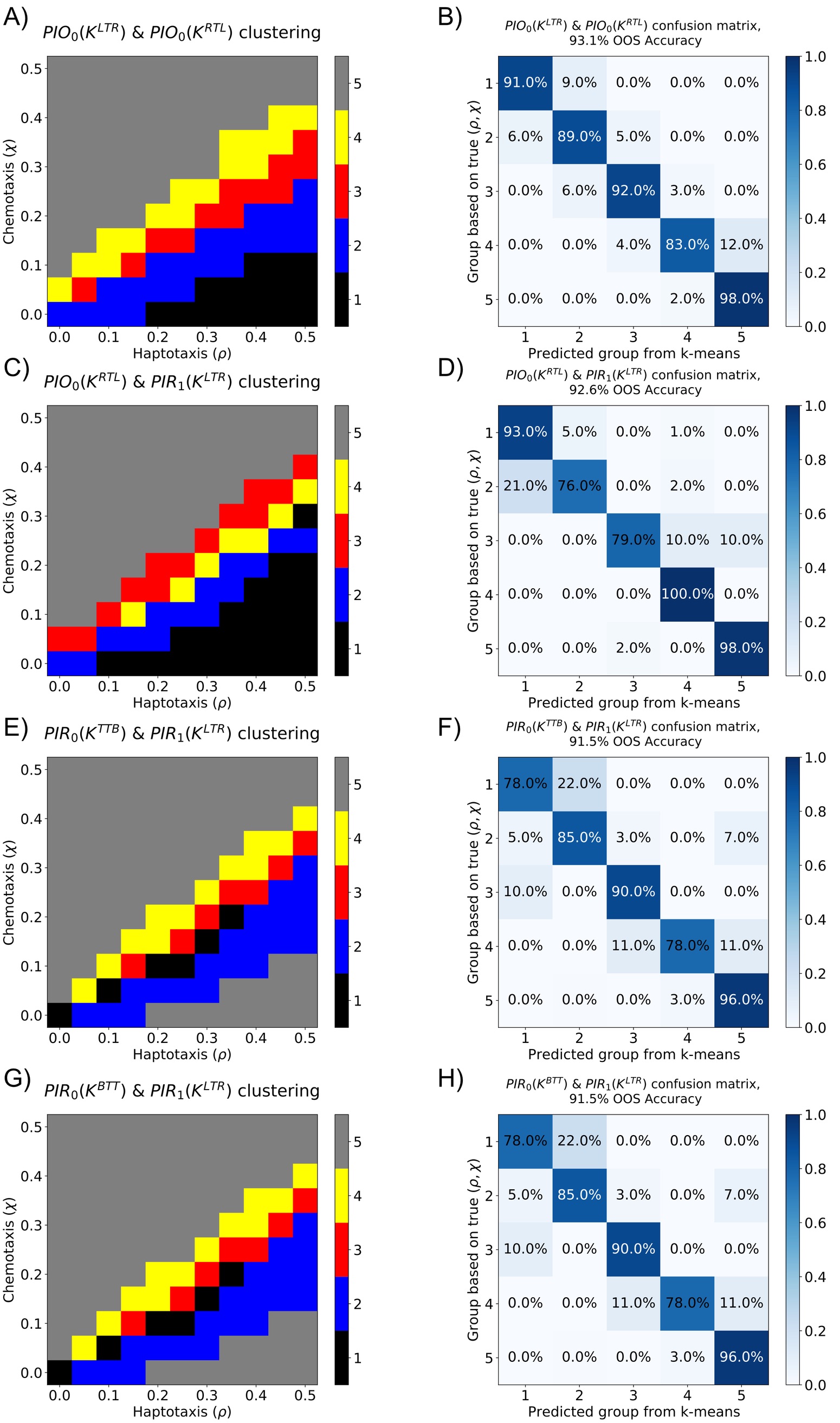}
    \caption{\textbf{Double plane sweeping descriptor vector clustering.} Clustering of the $(\rho,\chi)$ parameter space using $k$-means with $k=5$ on doubles of the sweeping plane topological filtration to summarize each simulated vasculature. The four highest OOS accuracies resulted from the A) $PIO_0(K^{LTR})$ \& $PIO_0(K^{RTL})$, C) $PIO_0(K^{RTL})$ \& $PIR_1(K^{LTR})$, E) $PIO_0(K^{TTB})$ \& $PIR_1(K^{LTR})$, and F) $PIO_0(K^{BTT})$ \& $PIR_1(K^{LTR})$ descriptor vectors. The five clusters are ordered according to the mean $\chi$ value within the cluster. Panels B,D,F,H depict the out of sample confusion matrices for each descriptor vector.}
    \label{fig:SIfig10}
\end{figure}

\begin{table}
\centering
\begin{tabular}{|c|c|}
\hline 
\textbf{Parameter} & \textbf{Dimensional value}\tabularnewline
\hline 
$L$ length of domain & 2 mm\tabularnewline
\hline 
$D_n$ endothelial cell diffusion & $10^{-10}\nicefrac{\text{cm}^{2}}{s}$\tabularnewline
\hline 
$\chi_0$ chemotactic response & $2600\nicefrac{\text{cm}^{2}}{(M\times s)}$\tabularnewline
\hline 
$\rho_0$ haptotactic response & $2600\nicefrac{\text{cm}^{2}}{(M\times s)}$\tabularnewline
\hline 
$f_{0}$ fibronectin concentration & $10^{-10}$ M\tabularnewline
\hline 
$D_{c}$ TAF diffusion & $2.9\times10^{-7}\nicefrac{\text{cm}^{2}}{s}$\tabularnewline
\hline 
$\omega$ Production of fibronectin & $1.35\times10^{-16}$M/s\tabularnewline
\hline 
$\mu$ uptake of fibronectin & $1\times10^{-5}$M/s\tabularnewline
\hline 
$\lambda$ Uptake of TAF & $2.7\times10^{-6}$M/s\tabularnewline
\hline 
\end{tabular}

\vspace{.1cm}
\textbf{S1 Table. Anderson-Chaplain model parameters.} List of mechanistic parameters for the Chaplain Anderson ABM from \cite{anderson_continuous_1998}. 
\label{tab:SItab1}
\end{table}

\begin{table}[]
    \centering
    \begin{tabular}{|c|c|}
    \hline 
    \textbf{Parameter} & \textbf{baseline value}\tabularnewline
    \hline 
    $D$  & 0.00035 \\
    \hline 
        $\chi$ & 0.38 \\
        \hline 
        $\rho$ &  0.34\\
        \hline 
        $\beta$ & 0.05 \\
        \hline 
        $\gamma$ &  0.1 \\
        \hline 
        $\epsilon_1$ &  0.45 \\
        \hline 
        $\epsilon_2$ &  0.45 \\
        \hline 
        $k$ &  0.75 \\
        \hline 
        
    \end{tabular}
    
    \vspace{0.1cm}
    \textbf{S2 Table. Anderson-Chaplain model nondimensionalized parameters.} Baseline nondimensionalized mechanistic parameters for the Chaplain-Anderson Model from \cite{anderson_continuous_1998}.

    \label{tab:SItab2}
\end{table}

\clearpage
\begin{table}
\centering
\def\arraystretch{1.5}
\begin{tabular}{|c|c|c|}
\hline
Feature & In Sample Accuracy & Out of Sample Accuracy \\ 
\hline
PIR$_1(K^{flood})$ & 71.9\% & 65.8\% \\
\hline
$\beta$$_1(K^{flood})$ & 72.6\% & 63.4\% \\
\hline
PIO$_1(K^{flood})$ & 71.1\% & 60.6\% \\
\hline
$\beta$$_0(K^{flood})$ & 66.2\% & 52.6\% \\
\hline
PIO$_0(K^{flood})$ & 60.0\% & 46.0\% \\
\hline
PIR$_0(K^{flood})$ & 62.1\% & 44.9\% \\
\hline
\end{tabular}

\vspace{0.1cm}
\textbf{S3 Table. Individual plane sweeping clustering.} Out of Sample Accuracy scores for individual feature vectors from the flooding filtration using $k$-means classification with $k=5$.
\label{tab:SItab3}
\end{table}

\clearpage
\begin{table}
\centering
\def\arraystretch{1.5}
\begin{tabular}{|c|c|c|}
\hline
Feature & In Sample Accuracy & Out of Sample Accuracy \\ 
\hline
PIO$_0(K^{flood})$ \& PIR$_1(K^{flood})$ & 73.2\% & 66.4\% \\
\hline
PIR$_0(K^{flood})$ \& PIR$_1(K^{flood})$ & 71.9\% & 65.6\% \\
\hline
BC$_0(K^{flood})$ \& BC$_1(K^{flood})$ & 74.6\% & 65.0\% \\
\hline
PIO$_0(K^{flood})$ \& BC$_1(K^{flood})$ & 72.6\% & 63.9\% \\
\hline
PIR$_1(K^{flood})$ \& BC$_1(K^{flood})$ & 72.6\% & 63.4\% \\
\hline
PIR$_0(K^{flood})$ \& BC$_1(K^{flood})$ & 72.6\% & 63.4\% \\
\hline
PIO$_1(K^{flood})$ \& BC$_1(K^{flood})$ & 72.7\% & 63.4\% \\
\hline
PIO$_0(K^{flood})$ \& PIO$_1(K^{flood})$ & 70.7\% & 62.5\% \\
\hline
PIO$_1(K^{flood})$ \& PIR$_0(K^{flood})$ & 71.2\% & 60.6\% \\
\hline
PIO$_1(K^{flood})$ \& PIR$_1(K^{flood})$ & 70.6\% & 59.8\% \\
\hline
PIO$_0(K^{flood})$ \& BC$_0(K^{flood})$ & 66.4\% & 53.4\% \\
\hline
PIO$_1(K^{flood})$ \& BC$_0(K^{flood})$ & 66.6\% & 52.1\% \\
\hline
PIO$_0(K^{flood})$ \& PIR$_0(K^{flood})$ & 62.1\% & 51.5\% \\
\hline
PIR$_1(K^{flood})$ \& BC$_0(K^{flood})$ & 66.1\% & 51.0\% \\
\hline
PIR$_0(K^{flood})$ \& BC$_0(K^{flood})$ & 66.1\% & 51.0\% \\
\hline
\end{tabular}
\vspace{.1cm}
\textbf{S4 Table. Double flooding clustering.} Out of Sample Accuracy scores for doubles of feature vectors from the flooding filtration using $k$-means classification with $k=5$.
\label{tab:SItab4}
\end{table}

\clearpage
\begin{table}[]
    \centering
    \def\arraystretch{1.5}
    \begin{tabular}{|c|c|c|}
\hline
Feature & In Sample Accuracy & Out of Sample Accuracy \\ 
\hline
PIR$_1(K^{LTR})$ & 92.3\% & 91.5\% \\
\hline
PIO$_0(K^{LTR})$ & 90.1\% & 83.5\% \\
\hline
PIO$_1(K^{RTL})$ & 82.4\% & 74.9\% \\
\hline
PIO$_1(K^{LTR})$ & 82.6\% & 74.7\% \\
\hline
$\beta$$_1(K^{LTR})$ & 82.3\% & 74.1\% \\
\hline
PIR$_1(K^{RTL})$ & 78.3\% & 73.3\% \\
\hline
PIO$_0(K^{RTL})$ & 82.9\% & 73.0\% \\
\hline
$\beta$$_1(K^{RTL})$ & 77.9\% & 72.7\% \\
\hline
PIR$_0(K^{RTL})$ & 78.0\% & 72.7\% \\
\hline
$\beta$$_0(K^{RTL})$ & 79.0\% & 70.8\% \\
\hline
$\beta$$_1(K^{TTB})$ & 74.1\% & 68.6\% \\
\hline
$\beta$$_0(K^{LTR})$ & 77.7\% & 66.1\% \\
\hline
$\beta$$_1(K^{BTT})$ & 71.4\% & 65.0\% \\
\hline
PIO$_1(K^{BTT})$ & 73.3\% & 64.5\% \\
\hline
PIO$_1(K^{TTB})$ & 73.9\% & 64.5\% \\
\hline
$\beta$$_0(K^{BTT})$ & 72.3\% & 61.4\% \\
\hline
PIR$_1(K^{BTT})$ & 69.3\% & 60.6\% \\
\hline
PIO$_0(K^{TTB})$ & 68.0\% & 60.1\% \\
\hline
PIR$_1(K^{TTB})$ & 70.1\% & 59.8\% \\
\hline
PIO$_0(K^{BTT})$ & 67.9\% & 56.5\% \\
\hline
PIR$_0(K^{LTR})$ & 67.1\% & 54.8\% \\
\hline
$\beta$$_0(K^{TTB})$ & 64.6\% & 50.1\% \\
\hline
PIR$_0(K^{BTT})$ & 55.6\% & 41.6\% \\
\hline
PIR$_0(K^{TTB})$ & 51.9\% & 36.6\% \\
\hline
\end{tabular}

\vspace{0.1cm}
\textbf{S5 Table. Individual plane sweeping clustering.} Out of Sample Accuracy scores for individual feature vectors from the sweeping plane filtration using $k$-means classification with $k=5$.
\label{fig:SItab5}
\end{table}

\end{document}